\documentclass{article} 
\usepackage{amsmath,amssymb,rotating,color}

\newdimen \myunit
\newdimen \myhsize
\newdimen \myvsize
\newcommand{\wire}[1]{\raisebox{3\myunit}[0cm][0cm]{\small #1}}
\newcommand{\smwidth}{8}
\newcommand{\smhalf}{4}
\newcommand{\qthicklines}{\linethickness{1.5\myunit}}

\newcommand{\qheight}{10}
\newcommand{\qtopheight}{5}
\newcommand{\qtopheightminusthreehalves}{3.5}
\newcommand{\qtopheightplustwo}{7}
\newcommand{\qbottomheight}{5}
\newcommand{\qbottomheightminusthreehalves}{3.5}
\newcommand{\qbottomheightplustwo}{7}

\definecolor{darkgreen}{rgb}{0,0.5,0}

\def\cn#1{
\begin{picture}(4,\qheight)(0,0)
  \put(2,\qtopheight){\circle*{2}}
  \put(0,\qtopheight){\line(1,0){4}}
\ifcase #1
    \put(2,0){\line(0,1){\qheight}}
\or \put(2,\qtopheight){\line(0,1){\qbottomheight}}
\or \put(2,\qtopheight){\line(0,-1){\qtopheight}}
\fi
\end{picture}
}

\def\ccn#1#2{
\begin{picture}(4,\qheight)(0,0)
  \put(0,\qtopheight){\line(1,0){4}}
\textcolor{#1}{
  \put(2,\qtopheight){\circle*{2}}
\ifcase #2
    \put(2,0){\line(0,1){\qheight}}
\or \put(2,\qtopheight){\line(0,1){\qbottomheight}}
\or \put(2,\qtopheight){\line(0,-1){\qtopheight}}
\fi}
\end{picture}
}

\def\nn#1{
\begin{picture}(4,\qheight)(0,0)
  \put(2,\qtopheight){\circle{3}}
  \put(0,\qtopheight){\line(1,0){.5}}
  \put(4,\qtopheight){\line(-1,0){.5}}
\ifcase #1
    \put(2,0){\line(0,1){\qtopheightminusthreehalves}}
    \put(2,\qheight){\line(0,-1){\qbottomheightminusthreehalves}}
\or \put(2,\qheight){\line(0,-1){\qbottomheightminusthreehalves}}
\or \put(2,0){\line(0,1){\qtopheightminusthreehalves}}
\fi
\end{picture}
}

\def\cnn#1#2{
\begin{picture}(4,\qheight)(0,0)
  \textcolor{#1}{\put(2,\qtopheight){\circle{3}}}
  \put(0,\qtopheight){\line(1,0){.5}}
  \put(4,\qtopheight){\line(-1,0){.5}}
\textcolor{#1}{
\ifcase #2
    \put(2,0){\line(0,1){\qtopheightminusthreehalves}}
    \put(2,\qheight){\line(0,-1){\qbottomheightminusthreehalves}}
\or \put(2,\qheight){\line(0,-1){\qbottomheightminusthreehalves}}
\or \put(2,0){\line(0,1){\qtopheightminusthreehalves}}
\fi}
\end{picture}
}

\def\xn#1{
\begin{picture}(4,\qheight)(0,0)
  \put(2,\qtopheight){\circle{4}}
  \put(0,\qtopheight){\line(1,0){4}}
\ifcase #1
    \put(2,0){\line(0,1){\qheight}}
\or \put(2,\qheight){\line(0,-1){\qbottomheightplustwo}}
\or \put(2,0){\line(0,1){\qtopheightplustwo}}
\or \put(2,\qtopheightplustwo){\line(0,-1){4}}
\fi
\end{picture}
}

\def\cxn#1#2{
\begin{picture}(4,\qheight)(0,0)
  \textcolor{#1}{
  \put(2,\qtopheight){\circle{4}}
  \put(0,\qtopheight){\line(1,0){4}}
\ifcase #2
    \put(2,0){\line(0,1){\qheight}}
\or \put(2,\qheight){\line(0,-1){\qbottomheightplustwo}}
\or \put(2,0){\line(0,1){\qtopheightplustwo}}
\or \put(2,\qtopheightplustwo){\line(0,-1){4}}
\fi}
\end{picture}
}

\newcommand{\sn}[1]{
\begin{picture}(4,\qheight)(0,0)
\ifcase #1
    \put(0,\qtopheight){\line(1,0){4}}
\or \put(2,0){\line(0,1){\qheight}}
\or \put(0,\qtopheight){\line(1,0){4}}
    \put(2,0){\line(0,1){\qheight}}
\or \put(0,\qtopheight){\line(1,0){4}}
    \multiput(2,1)(0,\qtopheight){2}{\line(0,1){\qtopheightminusthreehalves}}
\fi
\end{picture}
}

\newcommand{\csn}[2]{
\begin{picture}(4,\qheight)(0,0)
\ifcase #2
    \put(0,\qtopheight){\line(1,0){4}}
\or \textcolor{#1}{\put(2,0){\line(0,1){\qheight}}}
\or \put(0,\qtopheight){\line(1,0){4}}
    \textcolor{#1}{\put(2,0){\line(0,1){\qheight}}}
\fi
\end{picture}
}

\def\sm#1{
\begin{picture}(\smwidth,\qheight)(0,0)
\ifcase #1
    \put(0,\qtopheight){\line(1,0){\smwidth}}
\or \put(\smhalf,0){\line(0,1){\qheight}}
\or \put(0,\qtopheight){\line(1,0){\smwidth}}
    \put(\smhalf,0){\line(0,1){\qheight}}
\or \put(0,\qtopheight){\line(1,0){\smwidth}}
    \multiput(\smhalf,1)(0,\qtopheight){2}{\line(0,1){\qtopheightminusthreehalves}}
\fi
\end{picture}
}

\def\sx#1{
\begin{picture}(12,\qheight)(0,0)
\ifcase #1
    \put(0,\qtopheight){\line(1,0){12}}
\or \put(6,0){\line(0,1){\qheight}}
\or \put(0,\qtopheight){\line(1,0){12}}
    \put(6,0){\line(0,1){\qheight}}
\or \multiput(0,\qtopheight)(2.6,0){\qtopheight}{\line(1,0){1.6}}
    \put(1,0){\line(0,1){\qheight}}
    \qthicklines
    \put(11,0){\line(0,1){\qheight}}
    \thinlines
\or \multiput(0,\qtopheight)(2.6,0){\qtopheight}{\line(1,0){1.6}}
    \qthicklines
    \put(1,0){\line(0,1){\qheight}}
    \thinlines
    \put(11,0){\line(0,1){\qheight}}
\fi
\end{picture}
}

\def\dx#1{
\begin{picture}(12,\qheight)(0,0)
  \put(6,\qtopheight){\circle*{2}}
  \put(0,\qtopheight){\line(1,0){12}}
\ifcase #1
    \put(6,0){\line(0,1){\qheight}}
\or \put(6,\qtopheight){\line(0,1){\qbottomheight}}
\or \put(6,\qtopheight){\line(0,-1){\qtopheight}}
\or {}
\or \put(1,0){\line(0,1){\qheight}}
    \qthicklines
    \put(11,0){\line(0,1){\qheight}}
    \thinlines
\or \qthicklines
    \put(1,0){\line(0,1){\qheight}}
    \thinlines
    \put(11,0){\line(0,1){\qheight}}
\fi
\end{picture}
}

\def\ex#1{
\begin{picture}(12,\qheight)(0,0)
  \put(6,\qtopheight){\circle{3}}
  \put(0,\qtopheight){\line(1,0){4.5}}
  \put(12,\qtopheight){\line(-1,0){4.5}}
\ifcase #1
    \put(6,0){\line(0,1){\qtopheightminusthreehalves}}
    \put(6,\qheight){\line(0,-1){\qbottomheightminusthreehalves}}
\or \put(6,\qheight){\line(0,-1){\qbottomheightminusthreehalves}}
\or \put(6,0){\line(0,1){\qtopheightminusthreehalves}}
\or {}
\or \put(1,0){\line(0,1){\qheight}}
    \qthicklines
    \put(11,0){\line(0,1){\qheight}}
    \thinlines
\or \qthicklines
    \put(1,0){\line(0,1){\qheight}}
    \thinlines
    \put(11,0){\line(0,1){\qheight}}
\fi
\end{picture}
}

\def\nt#1{
\begin{picture}(12,\qheight)(0,0)
  \put(6,\qtopheight){\circle{4}}
  \put(0,\qtopheight){\line(1,0){12}}
\ifcase #1
    \put(6,0){\line(0,1){\qheight}}
\or \put(6,\qheight){\line(0,-1){\qbottomheightplustwo}}
\or \put(6,0){\line(0,1){\qtopheightplustwo}}
\or \put(6,\qtopheightplustwo){\line(0,-1){4}}
\or \put(1,0){\line(0,1){\qheight}}
    \qthicklines
    \put(11,0){\line(0,1){\qheight}}
    \thinlines
    \put(6,\qtopheightplustwo){\line(0,-1){4}}
\or \qthicklines
    \put(1,0){\line(0,1){\qheight}}
    \thinlines
    \put(11,0){\line(0,1){\qheight}}
    \put(6,\qtopheightplustwo){\line(0,-1){4}}
\fi
\end{picture}
}

\def\ox#1{
\begin{picture}(12,\qheight)(0,0)
  \put(6,\qtopheight){\circle{3}}
  \put(0,\qtopheight){\line(1,0){12}}
\ifcase #1 
    \put(6,0){\line(0,1){\qheight}}
\or \put(6,\qheight){\line(0,-1){6.5}}
\or \put(6,0){\line(0,1){6.5}}
\fi
\end{picture}
}

\newcommand{\ct}[1]{
\begin{picture}(12,\qheight)(0,0)
  \multiput(0,\qtopheight)(11,0){2}{\line(1,0){1}}
  \put(6,\qtopheight){\circle{\qheight}}
  \put(0,0){\vbox to \myvsize{\vfill
	\hbox to \myhsize{\hfill #1\hfill}\vfill}}
\end{picture}
}

\newcommand{\ti}[1]{
\begin{picture}(12,\qheight)(0,0)
  \multiput(1,0)(\qheight,0){2}{\line(0,1){\qheight}}
  \multiput(1,0)(0,\qheight){2}{\line(1,0){\qheight}}
  \multiput(0,\qtopheight)(11,0){2}{\line(1,0){1}}
  \put(0,0){\vbox to \myvsize{\vfill
	\hbox to \myhsize{\hfill #1\hfill}\vfill}}
\end{picture}
}

\newcommand{\tc}[1]{
\begin{picture}(12,\qheight)(0,0)
 \put(6,\qtopheight){\circle{\qheight}}
 \multiput(0,\qtopheight)(11,0){2}{\line(1,0){1}}
 \put(0,0){\vbox to \myvsize{\vfill
	\hbox to \myhsize{\hfill #1\hfill}\vfill}}
\end{picture}
}

\newcommand{\tb}[2]{
\begin{picture}(12,\qheight)(0,0)
  \put(1,0){\line(0,1){\qheight}}
  \qthicklines
  \put(11,0){\line(0,1){\qheight}}
  \thinlines
  \multiput(1,0)(\qheight,0){2}{\line(0,1){\qheight}}
  \multiput(0,\qtopheight)(11,0){2}{\line(1,0){1}}
  \put(0,0){\vbox to \myvsize{\vfill
	\hbox to \myhsize{\hfill #2\hfill}\vfill}}
\ifcase #1
{}
\or \put(1,0){\line(1,0){\qheight}}
\or \put(1,\qheight){\line(1,0){\qheight}}
\or \multiput(1,0)(0,\qheight){2}{\line(1,0){\qheight}}
\fi
\end{picture}
}

\newcommand{\tp}[2]{
\begin{picture}(12,\qheight)(0,0)
  \qthicklines
  \put(1,0){\line(0,1){\qheight}}
  \thinlines
  \put(11,0){\line(0,1){\qheight}}
  \multiput(0,\qtopheight)(11,0){2}{\line(1,0){1}}
  \put(0,0){\vbox to \myvsize{\vfill
	\hbox to \myhsize{\hfill #2\hfill}\vfill}}
\ifcase #1
{}
\or \put(1,0){\line(1,0){\qheight}}
\or \put(1,\qheight){\line(1,0){\qheight}}
\or \multiput(1,0)(0,\qheight){2}{\line(1,0){\qheight}}
\fi
\end{picture}
}

\newcommand{\place}[1]{\vbox to \myvsize{\vfill
	\hbox to \myhsize{\hfill #1\hfill}\vfill}}
\def\plac#1#2{\vbox to \myvsize{\vfill
	\hbox to #1\myhsize{#2\hfill}\vfill}}

\newcommand{\xor}{\oplus}
\newcommand{\maj}{\mathop{\rm MAJ}\nolimits}
\newcommand{\csop}{\mathbin{\circledast}}
\newcommand{\xoreq}{\mathbin{\xor\!\!=}}
\newcommand{\floor}[1]{\left\lfloor #1 \right\rfloor} 
\newcommand{\ceil}[1]{\left\lceil #1 \right\rceil} 
\newcommand{\Z}{\mathbb Z}
\newcommand{\caps}[1]{{\sc #1}}
\def\QCLA{\caps{qcla}}
\def\CLA{\caps{cla}}
\def\AND{\caps{and}}
\def\NOT{\caps{not}}
\def\CNOT{controlled-{\NOT}}
\def\XOR{\caps{xor}}

\title{A Logarithmic-Depth Quantum Carry-Lookahead Adder}
\author{Thomas G. Draper\thanks{6013 Pontiac Street, Berwyn Heights, MD 20740.  {\tt tgd@math.umd.edu}}
\and Samuel A.~Kutin\thanks{Center for Communications Research, 805 Bunn Drive, Princeton, NJ 08540. {\tt kutin@idaccr.org}}
\and Eric M.~Rains\thanks{Mathematics, University of California, Davis, 1 Shields Avenue, Davis, CA 95616--8633. {\tt rains@math.ucdavis.edu}}
\and Krysta M.~Svore\thanks{Department of Computer Science, Columbia University, 1214 Amsterdam Avenue, New York, NY 10027--7003.  {\tt kmsvore@cs.columbia.edu}}}

\begin{document}

\maketitle
\begin{abstract}
We present an efficient addition circuit, borrowing
techniques from the classical carry-lookahead arithmetic circuit.  Our
quantum carry-lookahead (\QCLA) adder accepts two $n$-bit numbers
and adds them in $O(\log n)$ depth using $O(n)$ ancillary qubits.  We
present both in-place and out-of-place versions, as well as versions
that add modulo $2^n$ and modulo $2^n - 1$.

Previously, the linear-depth ripple-carry addition circuit has been
the method of choice.  Our work reduces the cost of addition
dramatically with only a slight increase in the number of required
qubits.  The \QCLA\ adder can be used within current modular
multiplication circuits to reduce substantially the run-time of
Shor's algorithm. 
\end{abstract}

\section{Introduction}
\label{intro-sec}
With the advent of Shor's algorithms for prime factorization and the
discrete logarithm problem, it is necessary to design efficient
quantum arithmetic circuits. Previous quantum addition circuits
include the quantum ripple-carry adder of Vedral, Barenco, and
Ekert~\cite{VBE}, which has recently been
improved~\cite{ripple-carry-add}, and the transform
adder~\cite{draper-add}.  Both of these approaches have depth linear in
the number of input bits.  We present a new adder whose depth is
logarithmic in the number of input bits.  The circuit size, and the
number of ancillary qubits needed, are linear in the number of input
bits.

Our technique is derived from classical methods that perform in time
logarithmic in the number of input bits.  The classical carry-lookahead
({\CLA}) adder \cite{Harvard,Ofm63,WeiSmi56}
computes the carry bits in a tree-like structure, yielding a logarithmic-depth
circuit. We can exploit this same structure to design a
quantum {\CLA} (\QCLA) circuit to add two $n$-bit numbers in $O(\log n)$ depth.
The \QCLA\ adder works in $\Z$, and can be modified to add (mod~$2^n$) or
(mod~$2^{n} - 1$).

The theory of carry-lookahead addition has been known for fifty
years~\cite{Harvard}, and has appeared in circuit design
textbooks~\cite[pp.~158--161]{Mano},\cite[pp.~84--91]{Hwang}.  Why,
then, is this paper necessary?  What are the challenges of adapting
the {\CLA} technique to a quantum circuit?  There are several constraints
we have to consider:
\begin{itemize}
\item Reversibility:  We are limited to operations which do not
destroy information.
\item Erasure:  If we use scratch space, we must explicitly erase it.
We will not be able to take advantage of quantum interference if our
circuit leaves extra information in scratch registers.
\item Space-boundedness:  We wish to minimize the use of ancillae. 
\item Bounded fan-out:  At a given time, we can only use a wire as an
input to a single quantum gate.  To use multiple copies of a bit, we
must explicitly perform a fan-out operation, which increases the size
and depth of the circuit, and may increase the necessary space.
\end{itemize}

In Section~\ref{cla-sec}, we discuss the classical theory of
carry-lookahead addition; we then adapt this theory to the quantum
setting in Sections~\ref{reversible-carry-status-sec}
and~\ref{qcla-sec}.  Next, in Section~\ref{extensions-sec}, we discuss
various modified versions of the addition problem: how to add (mod
$2^n$) or (mod $2^n - 1$), how to compare, and how to take an incoming
carry bit as input.  The complexities of the various circuits are
summarized in Table~\ref{summary-table} on
page~\pageref{summary-table}.  Finally, we close with some thoughts on
future work.

\section{Preliminaries}
\label{prelim-sec}
We first describe our notation throughout this paper.  We then discuss
the classical carry-lookahead adder.

\subsection{Notation}
\label{notation-sec}
We write the binary expansion of a number $r$ as $r=r_{n-1} r_{n-2}\cdots r_0$, where $r_0$ is the low-order bit.

We generally represent negative numbers using
{\em two's-complement arithmetic\/}, in which the bitwise complement
$r'$ is equal to $-r - 1$.
In Section~\ref{mod-mersenne-sec}, we consider
{\em one's-complement arithmetic\/}, in which $r' = -r$.  Note that, in this
latter scheme, the all-zeros bit string and all-ones bit string both represent
zero, so we have to be careful when designing reversible one's-complement
arithmetic circuits.

In our circuit diagrams, time runs from left to right.  We use the
standard notation for quantum circuit operations: $\oplus$ for
negation, and $\bullet$ for a control.

In this paper, our circuits are composed of \NOT\ gates (also called
negations), controlled-\NOT\ (\CNOT) gates, and Toffoli gates.  A \CNOT\
gate has a single control qubit connected to a \NOT\ gate on the target
qubit.  A Toffoli gate has two qubits controlling the application of a
\NOT\ gate to the target qubit.  Hence, all of our circuits are
classical reversible circuits.

We will refer to the two inputs to our addition circuit as $a$ and $b$.
Our goal is to compute the sum $s$, either in place (on top of $b$) or
out of place.  We compute $s$ by first finding $c$, the {\em carry bits\/},
such that $s = a \xor b \xor c$.  (If one computes the sum using standard
school-book addition, then $c$ is the sequence of carries.)

We let $w(n)$ denote the number of ones in the binary expansion of $n$.
We observe that
\begin{equation}
\label{wn-eqn}
n - w(n) = \sum_{i=1}^\infty \floor{n \over 2^i}.
\end{equation}

We denote $\log_2$ simply by $\log$.

\subsection{Classical carry-lookahead addition}
\label{cla-sec}
In this section, we describe the classical carry-lookahead addition circuit
and our motivation for using a carry-lookahead structure. The {\CLA} adder
\cite{Ofm63} sums two $n$-bit numbers in $O(\log n)$ depth.  In this
arithmetic circuit,
partial information about the incoming carry bits is exploited to avoid a
linear-time ripple-carry computation.  The carry bit string can be computed
using a tree structure to greatly reduce the number of required operations.
 
The key ingredient is the {\em carry status\/} on an
interval, denoted $C[i,j]$.  This status can take one of three
values:  {\tt k} represents ``kill,'' {\tt g} represents ``generate,''
and {\tt p} represents ``propagate.''  We begin with a discussion of
the carry status $C[i,i+1]$.

Suppose we are adding $a$ and $b$, and we have computed the carry bit
$c_i$.  The next carry bit $c_{i+1}$ is the majority function
$\maj(a_i,b_i,c_i)$.  The base case for this process, $c_0$, is
assumed to be $0$---see Section~\ref{incoming-carry-sec} for
discussion of the more general problem where $c_0$ is an input bit.

When $a_i = b_i$, we can determine the carry bit $c_{i+1}$ without
knowing $c_i$.  Specifically, if $a_i=b_i=0$, then the outgoing carry
bit $c_{i+1}$ is automatically ``killed'' and we set $c_{i+1}=0$; we
say that $C[i,i+1] = {\tt k}$.  Similarly, if $a_i = b_i =1$, then a
carry bit is ``generated'' and $c_{i+1}=1$ with carry status $C[i,i+1]
= {\tt g}$.

If $a_i\neq b_i$, then we cannot determine $c_{i+1}$ without
knowing $c_i$.  In this case, the carry $c_i$ is ``propagated''
and we set $c_{i+1}=c_i$ with carry status $C[i,i+1] = {\tt p}$.
Figure~\ref{status} summarizes this computation of the carry status.

\begin{figure}[h]
\begin{center}
\begin{tabular}{cc|cc}
$a_{i}$&$b_{i}$&$c_{i+1}$&$C[i,i+1]$\\ \hline
0&0&0&k\\
0&1&$c_{i}$&{\tt p}\\
1&0&$c_{i}$&{\tt p}\\
1&1&1&{\tt g}\\
\end{tabular}
\caption{The carry status of $a_i$ and $b_i$.}\label{status}
\end{center}
\end{figure}

Given $C[i-1, i]$ and $C[i,i+1]$, we can compute the carry status
$C[i-1,i+1]$.  The calculation is shown in Figure~\ref{FA}; we use
$\csop$ to denote the carry status operator.  If
$C[i-1,i+1] = {\tt k}$, then either a carry is killed at position $i$,
or it would be propagated at position $i$ but has been killed at position $i-1$.
Either way, if $C[i-1,i+1] = {\tt k}$, we know that $c_{i+1} = 0$.  Similarly,
if $C[i-1,i+1] = {\tt g}$, we know that $c_{i+1} = 1$.  If $C[i-1,i+1] = {\tt p}$,
we conclude that $c_{i+1} = c_{i-1}$.

\begin{figure}
\begin{center}
\begin{tabular}{cc|ccc}
&&&{$C[i,i+1]$}&\\
&$\csop$&{\tt k}&{\tt p}&{\tt g}\\ \hline
&{\tt k}&{\tt k}&{\tt k}&{\tt g}\\ 
$C[i-1,i]$&{\tt p}&{\tt k}&{\tt p}&{\tt g}\\
&{\tt g}&{\tt k}&{\tt g}&{\tt g}\\
\end{tabular}\caption{The carry status assignments $C[i-1,i+1]$
given the previous and current carry status values.}\label{FA}
\end{center}
\end{figure}

The carry status operator $\csop$ shown in Figure~\ref{FA} allows us to
merge intervals: for any $k$ satisfying $i < k < j$,
$$
C[i,j] = C[i,k] \csop C[k,j].
$$
The choice of $k$ does not affect the answer, since $\csop$ is
associative.  By successively doubling the sizes of intervals, we can
use this approach to compute $C[i,j]$ for any $i,j$ in logarithmic
depth.

We now describe the computation of the carry bits in detail.  Since
$C[i,j]$ can take three values, we must specify an encoding of $C[i,j]$
in bits.  We define $p[i,j]$ to be 1 when $C[i,j] = {\tt p}$, and we
define $g[i,j]$ to be 1 when $C[i,j] = {\tt g}$.  The relationship between
$C[i,j]$, $p[i,j]$, and $g[i,j]$ is depicted in Figure~\ref{cpg-table}.
Note that, in particular, we never have $p[i,j]=g[i,j]=1$.

\begin{figure}[h]
\begin{center}
\begin{tabular}{r|cc}
$C[i,j]$ & $p[i,j]$ & $g[i,j]$ \\ \hline
{\tt k} & 0 & 0 \\
{\tt g} & 0 & 1 \\
{\tt p} & 1 & 0 \\
\end{tabular}\caption{The carry status $C[i,j]$ encoded in two
bits $p[i,j]$ and $g[i,j]$.}\label{cpg-table}
\end{center}
\end{figure}

For any $i,j$, $p[i,j]$ is 1 if a carry propagates from bit position
$i$ to bit position $j$, and 0 otherwise.  Note that this occurs if
and only if $a_\ell \oplus b_\ell=1$ whenever $i \le \ell < j$.  For
any $k$ between $i$ and $j$, a carry bit is propagated from bit $i$ to
bit $j$ if a carry bit is propagated from $i$ to bit $k$, and then
also propagated from bit $k$ to bit $j$.  Thus, the computation of the
propagate bits, for any $i<k <j$, is
\begin{equation}
\label{pij-eqn}
p[i,j]=p[i,k] \wedge p[k,j].
\end{equation}

Next, we consider $g[i,j]$.  This quantity is 1 when a carry is generated
between bit positions $i$ and $j$.
The computation of the generate bits, for $i<k < j$, is
\begin{align}
\notag g[i,j] & = g[k,j] \vee (g[i,k] \wedge p[k,j]) \\
\label{gij-eqn} & = g[k,j] \xor (g[i,k] \wedge p[k,j]).
\end{align}
That is, either a carry bit is generated between bits $k$ and $j$, or
a carry bit is generated between bits $i$ and $k$ and then propagated
from bit $k$ to bit $j$. The second expression follows from the
observation that $g[k,j]$ and $p[k,j]$ cannot both be equal to 1.

For all $j > 0$, $p[0,j]$ is $0$, and $g[0,j]$ is the carry bit $c_j$.
By successively doubling the sizes of the intervals under consideration,
we can compute all carry bits in logarithmic depth.
%
%
%
%
%

\section{Reversible computation of carry status}
\label{reversible-carry-status-sec}

We are now ready to build a quantum adder using the {\CLA} technique.
We first explain how we can compute the carry status reversibly.

The circuit of this section has two input arrays, each of length
$n$:  $P_0$, initialized to $P_0[i] = p[i,i+1]$, and $G$,
initialized to $G[i] = g[i-1,i]$.  Note that the array $P_0$ is $0$-based,
but the array $G$ is $1$-based.  We also use $n - w(n) - \floor{\log n}$
ancillary bits, initialized to zero.

At the end of the computation, we want $G[i] = g[0,i] = c_i$.
We also need to erase our scratch work:  we must ensure that,
when we're done,  $P_0[i] = p[i,i+1]$
and the ancillary bits are reset to zero.

We will have roughly $\floor{\log n}$ rounds each of four different types:
\begin{enumerate}
\item $P$-rounds:  Compute $p[i,j]$ values into the ancillary space.
\item $G$-rounds:  Set $G[j] = g[i,j]$; for each $j$, we choose a
particular $i$ value.\footnote{We have $i = (j-1) \wedge j$,
where $-$ denotes subtraction in $\Z$ and $\wedge$ denotes bitwise {\AND}.}
\item $C$-rounds:  Set $G[j] = c_{j}$.
\item $P^{-1}$-rounds:  Erase the work done in the $P$-rounds.
\end{enumerate}
We first describe
the sequence of gates, and then we compute the circuit depth.

In $P$-round $t$, we compute all $p$-values of the form $p[i, j]$
where $i = 2^t m$ and $j = i + 2^t$.  We refer to these values
as $P_t[m]$, for $1 \le m < \floor{n/2^t}$.  We store these
$\floor{n/2^t} - 1$ values in our ancillary space.  By~(\ref{wn-eqn}),
the total space needed for all of the $P$-rounds is
$n - w(n) - \floor{\log n}$ bits.  We do not need to compute values
of the form $p[0, 2^t]$, since no carry is generated at 0; in
particular, when $t = \floor{\log n}$, no computation is done.

We compute $p[i,j]$ using~(\ref{pij-eqn}), with $k = 2^t m + 2^{t-1}$.
Note that both $p[i,k]$ and $p[k,j]$ were computed in $P$-round $(t-1)$,
so we can write $p[i,j]$ to the appropriate location using one Toffoli
gate.  The total number of gates is thus $n - w(n) - \floor{\log n}$.

In $G$-round $t$, we compute all $g$-values of the form $g[i,j]$
where $i = 2^t m$ and $j = i + 2^t$.  We store this value in
the location $G[j]$.  We use~(\ref{gij-eqn}), with $k = 2^t m + 2^{t-1}$.
Since $g[k,j]$ is already in location
$G[j]$ after $G$-round $(t-1)$, we can do this computation with a
single Toffoli gate, combining $g[i,k]$ (computed in $G$-round $(t-1)$)
and $p[k,j]$ (computed in $P$-round $(t-1)$).  The total number
of gates is $n - w(n)$.

In $C$-round $t$, we compute all $g$-values of the
form $g[0,j]$ with $j = 2^t m + 2^{t-1}$.  We begin with the
maximum $t$ for which some $j$ exists, $t = \floor{\log{2n \over 3}}
= 1 + \floor{\log{n \over 3}}$,
and work our way down to $t = 1$.
We use~(\ref{gij-eqn}) with $k = 2^t m$.  Since $g[k,j]$ is
already in location $G[j]$, we again need just one Toffoli gate:
we require $p[k,j]$ (computed in $P$-round $(t-1)$) and $g[0,k]$
(computed in $C$-round $(t+1)$ or earlier).  The total number of
gates is $n - \floor{\log n} - 1$.

Finally, in the $P^{-1}$-rounds, we simply repeat the same Toffolis
as in the $P$-rounds, in reverse order.

In summary, we must perform the following steps:
\begin{enumerate}
\item $P$-rounds.
For $t = 1$ to $\floor{\log n} - 1$:  for $1 \le m < \floor{n/2^t}$:
$$
P_t[m] \xoreq P_{t-1}[2m] P_{t-1}[2m+1].
$$
\item $G$-rounds.
For $t = 1$ to $\floor{\log n}$:  for $0 \le m < \floor{n/2^t}$:
$$
G[2^t m + 2^t] \xoreq G[2^t m + 2^{t-1}] P_{t-1}[2m+1].
$$

\item $C$-rounds.
For $t = \floor{\log{2n \over 3}}$ down to $1$:  for $1 \le m \le \floor{(n - 2^{t-1})/2^t}$:
$$
G[2^t m + 2^{t-1}] \xoreq G[2^t m] P_{t-1}[2m].
$$

\item $P^{-1}$-rounds.
For $t = \floor{\log n} - 1$ down to $1$:  for $1 \le m < \floor{n/2^t}$:
$$
P_t[m] \xoreq P_{t-1}[2m] P_{t-1}[2m+1].
$$
\end{enumerate}

The circuit consists of
\begin{equation}
\label{oneway-size}
4n - 3w(n) - 3 \floor{\log n} - 1
\end{equation}
Toffoli gates.

It would seem that the circuit described above would require
roughly $4 \log n$ time-slices.  However, we can overlap some of
the computation.

We start with $P$-round 1, which uses the arrays $P_0$ and $P_1$.
Then, $P$-round 2 uses the arrays $P_1$ and $P_2$.  Note that
$G$-round 1 uses the arrays $G$ and $P_0$; hence, we can run
$G$-round 1 in the same time-slice as $P$-round 2.  In general, we
can run $P$-round $(t+1)$ and $G$-round $t$ in parallel.

Similarly, once we have run $C$-round $t$, we are done using
$P_{t-1}$.  While we run $C$-round $(t-1)$, we can run $P^{-1}$-round $t$,
which uses $P_{t-1}$ to erase $P_t$.  We run $C$-round 1 in parallel
with $P$-round 2; we then need one additional time-slice to run
$P$-round 1.

So, the circuit has a depth of
\begin{equation}
\label{oneway-depth}
\floor{\log n} + \floor{\log{n \over 3}} + 3.
\end{equation}
For $n \le 3$,
expression~(\ref{oneway-depth}) overcounts the depth, since there
are no $P$-rounds.

\section{The complete quantum addition circuit}
\label{qcla-sec}

We are now ready to describe our quantum carry-lookahead addition circuit.
We first discuss the out-of-place version in Section~\ref{oop-qcla-sec},
and then the in-place version in Section~\ref{ip-qcla-sec}.

The out-of-place version produces $n+1$ bits of output, and uses
$n - w(n) - \floor{\log n}$ ancillae.  The depth is $2 \log n + O(1)$ and the
size is $8 n - O(\log n)$ gates.

The in-place version produces $1$ bit of output, and uses
$2n - w(n) - \floor{\log n} - 1$ ancillae.  The depth is
$4 \log n + O(1)$ and the size is $16n - O(\log n)$ gates.

Table~\ref{summary-table} on page~\pageref{summary-table}
summarizes the complexities of these two
adders, as well as the variants discussed in Section~\ref{extensions-sec}.

\subsection{Addition out of place}
\label{oop-qcla-sec}

We would like to add two $n$-bit numbers, $a$ and $b$, stored in
arrays $A$ and $B$.  We need
$n+1$ bits for the output, denoted by $Z$, and $n - w(n) - \floor{\log n}$
ancillary bits, denoted by $X$.  We assume that $Z$ and $X$ are initialized
to zero.  In the end, we want $Z$ to contain the quantity $s = a+b$.

\begin{figure}[p]
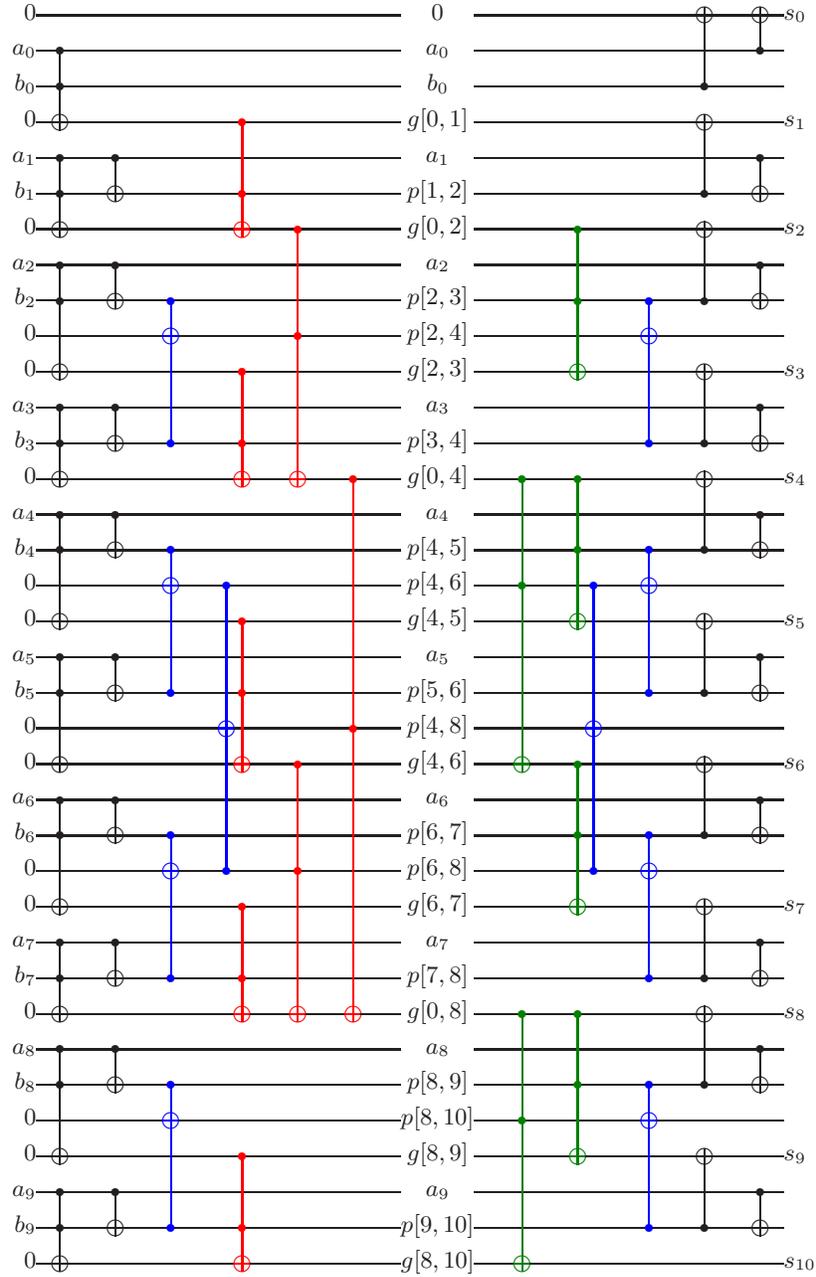

\begin{center}
\renewcommand{\arraystretch}{0}
\renewcommand{\smwidth}{10}
\renewcommand{\smhalf}{5}
\renewcommand{\qheight}{9}
\renewcommand{\qtopheight}{4}
\renewcommand{\qtopheightminusthreehalves}{2.5}
\renewcommand{\qtopheightplustwo}{6}
\renewcommand{\qbottomheight}{5}
\renewcommand{\qbottomheightminusthreehalves}{3.5}
\renewcommand{\qbottomheightplustwo}{7}
\begin{tabular}{r@{}*{27}{c@{}}l}
\wire{$0$} &\sn0&\sn0&\sm0&\sn0&\sm0&\sn0&\sm0&\sn0&\sn0&\sm0&\sn0&\sm0&\sn0&\sm0& \wire{$0$} &\sm0&\sn0&\sm0&\sn0&\sn0&\sm0&\sn0&\sm0&\xn2&\sm0&\xn2&\sn0& \wire{$s_0$} \\
\wire{$a_0$} &\sn0&\cn2&\sm0&\sn0&\sm0&\sn0&\sm0&\sn0&\sn0&\sm0&\sn0&\sm0&\sn0&\sm0& \wire{$a_0$} &\sm0&\sn0&\sm0&\sn0&\sn0&\sm0&\sn0&\sm0&\sn2&\sm0&\cn1&\sn0& \\
\wire{$b_0$} &\sn0&\cn0&\sm0&\sn0&\sm0&\sn0&\sm0&\sn0&\sn0&\sm0&\sn0&\sm0&\sn0&\sm0& \wire{$b_0$} &\sm0&\sn0&\sm0&\sn0&\sn0&\sm0&\sn0&\sm0&\cn1&\sm0&\sn0&\sn0& \\
\wire{$0$} &\sn0&\xn1&\sm0&\sn0&\sm0&\sn0&\sm0&\sn0&\ccn{red}2&\sm0&\sn0&\sm0&\sn0&\sm0& \wire{$g[0,1]$} &\sm0&\sn0&\sm0&\sn0&\sn0&\sm0&\sn0&\sm0&\xn2&\sm0&\sn0&\sn0& \wire{$s_1$} \\
\wire{$a_1$} &\sn0&\cn2&\sm0&\cn2&\sm0&\sn0&\sm0&\sn0&\csn{red}2&\sm0&\sn0&\sm0&\sn0&\sm0& \wire{$a_1$} &\sm0&\sn0&\sm0&\sn0&\sn0&\sm0&\sn0&\sm0&\sn2&\sm0&\cn2&\sn0& \\
\wire{$b_1$} &\sn0&\cn0&\sm0&\xn1&\sm0&\sn0&\sm0&\sn0&\ccn{red}0&\sm0&\sn0&\sm0&\sn0&\sm0& \wire{$p[1,2]$} &\sm0&\sn0&\sm0&\sn0&\sn0&\sm0&\sn0&\sm0&\cn1&\sm0&\xn1&\sn0& \\
\wire{$0$} &\sn0&\xn1&\sm0&\sn0&\sm0&\sn0&\sm0&\sn0&\cxn{red}1&\sm0&\ccn{red}2&\sm0&\sn0&\sm0& \wire{$g[0,2]$} &\sm0&\sn0&\sm0&\ccn{darkgreen}2&\sn0&\sm0&\sn0&\sm0&\xn2&\sm0&\sn0&\sn0& \wire{$s_2$} \\
\wire{$a_2$} &\sn0&\cn2&\sm0&\cn2&\sm0&\sn0&\sm0&\sn0&\sn0&\sm0&\csn{red}2&\sm0&\sn0&\sm0& \wire{$a_2$} &\sm0&\sn0&\sm0&\csn{darkgreen}2&\sn0&\sm0&\sn0&\sm0&\sn2&\sm0&\cn2&\sn0& \\
\wire{$b_2$} &\sn0&\cn0&\sm0&\xn1&\sm0&\ccn{blue}2&\sm0&\sn0&\sn0&\sm0&\csn{red}2&\sm0&\sn0&\sm0& \wire{$p[2,3]$} &\sm0&\sn0&\sm0&\ccn{darkgreen}0&\sn0&\sm0&\ccn{blue}2&\sm0&\cn1&\sm0&\xn1&\sn0& \\
\wire{$0$} &\sn0&\sn2&\sm0&\sn0&\sm0&\cxn{blue}0&\sm0&\sn0&\sn0&\sm0&\ccn{red}0&\sm0&\sn0&\sm0& \wire{$p[2,4]$} &\sm0&\sn0&\sm0&\csn{darkgreen}2&\sn0&\sm0&\cxn{blue}0&\sm0&\sn0&\sm0&\sn0&\sn0& \\
\wire{$0$} &\sn0&\xn1&\sm0&\sn0&\sm0&\csn{blue}2&\sm0&\sn0&\ccn{red}2&\sm0&\csn{red}2&\sm0&\sn0&\sm0& \wire{$g[2,3]$} &\sm0&\sn0&\sm0&\cxn{darkgreen}1&\sn0&\sm0&\csn{blue}2&\sm0&\xn2&\sm0&\sn0&\sn0& \wire{$s_3$} \\
\wire{$a_3$} &\sn0&\cn2&\sm0&\cn2&\sm0&\csn{blue}2&\sm0&\sn0&\csn{red}2&\sm0&\csn{red}2&\sm0&\sn0&\sm0& \wire{$a_3$} &\sm0&\sn0&\sm0&\sn0&\sn0&\sm0&\csn{blue}2&\sm0&\sn2&\sm0&\cn2&\sn0& \\
\wire{$b_3$} &\sn0&\cn0&\sm0&\xn1&\sm0&\ccn{blue}1&\sm0&\sn0&\ccn{red}0&\sm0&\csn{red}2&\sm0&\sn0&\sm0& \wire{$p[3,4]$} &\sm0&\sn0&\sm0&\sn0&\sn0&\sm0&\ccn{blue}1&\sm0&\cn1&\sm0&\xn1&\sn0& \\
\wire{$0$} &\sn0&\xn1&\sm0&\sn0&\sm0&\sn0&\sm0&\sn0&\cxn{red}1&\sm0&\cxn{red}1&\sm0&\ccn{red}2&\sm0& \wire{$g[0,4]$} &\sm0&\ccn{darkgreen}2&\sm0&\ccn{darkgreen}2&\sn0&\sm0&\sn0&\sm0&\xn2&\sm0&\sn0&\sn0& \wire{$s_4$} \\
\wire{$a_4$} &\sn0&\cn2&\sm0&\cn2&\sm0&\sn0&\sm0&\sn0&\sn0&\sm0&\sn0&\sm0&\csn{red}2&\sm0& \wire{$a_4$} &\sm0&\csn{darkgreen}2&\sm0&\csn{darkgreen}2&\sn0&\sm0&\sn0&\sm0&\sn2&\sm0&\cn2&\sn0& \\
\wire{$b_4$} &\sn0&\cn0&\sm0&\xn1&\sm0&\ccn{blue}2&\sm0&\sn0&\sn0&\sm0&\sn0&\sm0&\csn{red}2&\sm0& \wire{$p[4,5]$} &\sm0&\csn{darkgreen}2&\sm0&\ccn{darkgreen}0&\sn0&\sm0&\ccn{blue}2&\sm0&\cn1&\sm0&\xn1&\sn0& \\
\wire{$0$} &\sn0&\sn2&\sm0&\sn0&\sm0&\cxn{blue}0&\sm0&\ccn{blue}2&\sn0&\sm0&\sn0&\sm0&\csn{red}2&\sm0& \wire{$p[4,6]$} &\sm0&\ccn{darkgreen}0&\sm0&\csn{darkgreen}2&\ccn{blue}2&\sm0&\cxn{blue}0&\sm0&\sn0&\sm0&\sn0&\sn0& \\
\wire{$0$} &\sn0&\xn1&\sm0&\sn0&\sm0&\csn{blue}2&\sm0&\csn{blue}2&\ccn{red}2&\sm0&\sn0&\sm0&\csn{red}2&\sm0& \wire{$g[4,5]$} &\sm0&\csn{darkgreen}2&\sm0&\cxn{darkgreen}1&\csn{blue}2&\sm0&\csn{blue}2&\sm0&\xn2&\sm0&\sn0&\sn0& \wire{$s_5$} \\
\wire{$a_5$} &\sn0&\cn2&\sm0&\cn2&\sm0&\csn{blue}2&\sm0&\csn{blue}2&\csn{red}2&\sm0&\sn0&\sm0&\csn{red}2&\sm0& \wire{$a_5$} &\sm0&\csn{darkgreen}2&\sm0&\sn0&\csn{blue}2&\sm0&\csn{blue}2&\sm0&\sn2&\sm0&\cn2&\sn0& \\
\wire{$b_5$} &\sn0&\cn0&\sm0&\xn1&\sm0&\ccn{blue}1&\sm0&\csn{blue}2&\ccn{red}0&\sm0&\sn0&\sm0&\csn{red}2&\sm0& \wire{$p[5,6]$} &\sm0&\csn{darkgreen}2&\sm0&\sn0&\csn{blue}2&\sm0&\ccn{blue}1&\sm0&\cn1&\sm0&\xn1&\sn0& \\
\wire{$0$} &\sn0&\sn2&\sm0&\sn0&\sm0&\sn0&\sm0&\cxn{blue}0&\csn{red}2&\sm0&\sn0&\sm0&\ccn{red}0&\sm0& \wire{$p[4,8]$} &\sm0&\csn{darkgreen}2&\sm0&\sn0&\cxn{blue}0&\sm0&\sn0&\sm0&\sn0&\sm0&\sn0&\sn0& \\
\wire{$0$} &\sn0&\xn1&\sm0&\sn0&\sm0&\sn0&\sm0&\csn{blue}2&\cxn{red}1&\sm0&\ccn{red}2&\sm0&\csn{red}2&\sm0& \wire{$g[4,6]$} &\sm0&\cxn{darkgreen}1&\sm0&\ccn{darkgreen}2&\csn{blue}2&\sm0&\sn0&\sm0&\xn2&\sm0&\sn0&\sn0& \wire{$s_6$} \\
\wire{$a_6$} &\sn0&\cn2&\sm0&\cn2&\sm0&\sn0&\sm0&\csn{blue}2&\sn0&\sm0&\csn{red}2&\sm0&\csn{red}2&\sm0& \wire{$a_6$} &\sm0&\sn0&\sm0&\csn{darkgreen}2&\csn{blue}2&\sm0&\sn0&\sm0&\sn2&\sm0&\cn2&\sn0& \\
\wire{$b_6$} &\sn0&\cn0&\sm0&\xn1&\sm0&\ccn{blue}2&\sm0&\csn{blue}2&\sn0&\sm0&\csn{red}2&\sm0&\csn{red}2&\sm0& \wire{$p[6,7]$} &\sm0&\sn0&\sm0&\ccn{darkgreen}0&\csn{blue}2&\sm0&\ccn{blue}2&\sm0&\cn1&\sm0&\xn1&\sn0& \\
\wire{$0$} &\sn0&\sn2&\sm0&\sn0&\sm0&\cxn{blue}0&\sm0&\ccn{blue}1&\sn0&\sm0&\ccn{red}0&\sm0&\csn{red}2&\sm0& \wire{$p[6,8]$} &\sm0&\sn0&\sm0&\csn{darkgreen}2&\ccn{blue}1&\sm0&\cxn{blue}0&\sm0&\sn0&\sm0&\sn0&\sn0& \\
\wire{$0$} &\sn0&\xn1&\sm0&\sn0&\sm0&\csn{blue}2&\sm0&\sn0&\ccn{red}2&\sm0&\csn{red}2&\sm0&\csn{red}2&\sm0& \wire{$g[6,7]$} &\sm0&\sn0&\sm0&\cxn{darkgreen}1&\sn0&\sm0&\csn{blue}2&\sm0&\xn2&\sm0&\sn0&\sn0& \wire{$s_7$} \\
\wire{$a_7$} &\sn0&\cn2&\sm0&\cn2&\sm0&\csn{blue}2&\sm0&\sn0&\csn{red}2&\sm0&\csn{red}2&\sm0&\csn{red}2&\sm0& \wire{$a_7$} &\sm0&\sn0&\sm0&\sn0&\sn0&\sm0&\csn{blue}2&\sm0&\sn2&\sm0&\cn2&\sn0& \\
\wire{$b_7$} &\sn0&\cn0&\sm0&\xn1&\sm0&\ccn{blue}1&\sm0&\sn0&\ccn{red}0&\sm0&\csn{red}2&\sm0&\csn{red}2&\sm0& \wire{$p[7,8]$} &\sm0&\sn0&\sm0&\sn0&\sn0&\sm0&\ccn{blue}1&\sm0&\cn1&\sm0&\xn1&\sn0& \\
\wire{$0$} &\sn0&\xn1&\sm0&\sn0&\sm0&\sn0&\sm0&\sn0&\cxn{red}1&\sm0&\cxn{red}1&\sm0&\cxn{red}1&\sm0& \wire{$g[0,8]$} &\sm0&\ccn{darkgreen}2&\sm0&\ccn{darkgreen}2&\sn0&\sm0&\sn0&\sm0&\xn2&\sm0&\sn0&\sn0& \wire{$s_8$} \\
\wire{$a_8$} &\sn0&\cn2&\sm0&\cn2&\sm0&\sn0&\sm0&\sn0&\sn0&\sm0&\sn0&\sm0&\sn0&\sm0& \wire{$a_8$} &\sm0&\csn{darkgreen}2&\sm0&\csn{darkgreen}2&\sn0&\sm0&\sn0&\sm0&\sn2&\sm0&\cn2&\sn0& \\
\wire{$b_8$} &\sn0&\cn0&\sm0&\xn1&\sm0&\ccn{blue}2&\sm0&\sn0&\sn0&\sm0&\sn0&\sm0&\sn0&\sm0& \wire{$p[8,9]$} &\sm0&\csn{darkgreen}2&\sm0&\ccn{darkgreen}0&\sn0&\sm0&\ccn{blue}2&\sm0&\cn1&\sm0&\xn1&\sn0& \\
\wire{$0$} &\sn0&\sn2&\sm0&\sn0&\sm0&\cxn{blue}0&\sm0&\sn0&\sn0&\sm0&\sn0&\sm0&\sn0&\sm0& \wire{$p[8,10]$} &\sm0&\ccn{darkgreen}0&\sm0&\csn{darkgreen}2&\sn0&\sm0&\cxn{blue}0&\sm0&\sn0&\sm0&\sn0&\sn0& \\
\wire{$0$} &\sn0&\xn1&\sm0&\sn0&\sm0&\csn{blue}2&\sm0&\sn0&\ccn{red}2&\sm0&\sn0&\sm0&\sn0&\sm0& \wire{$g[8,9]$} &\sm0&\csn{darkgreen}2&\sm0&\cxn{darkgreen}1&\sn0&\sm0&\csn{blue}2&\sm0&\xn2&\sm0&\sn0&\sn0& \wire{$s_9$} \\
\wire{$a_9$} &\sn0&\cn2&\sm0&\cn2&\sm0&\csn{blue}2&\sm0&\sn0&\csn{red}2&\sm0&\sn0&\sm0&\sn0&\sm0& \wire{$a_9$} &\sm0&\csn{darkgreen}2&\sm0&\sn0&\sn0&\sm0&\csn{blue}2&\sm0&\sn2&\sm0&\cn2&\sn0& \\
\wire{$b_9$} &\sn0&\cn0&\sm0&\xn1&\sm0&\ccn{blue}1&\sm0&\sn0&\ccn{red}0&\sm0&\sn0&\sm0&\sn0&\sm0& \wire{$p[9,10]$} &\sm0&\csn{darkgreen}2&\sm0&\sn0&\sn0&\sm0&\ccn{blue}1&\sm0&\cn1&\sm0&\xn1&\sn0& \\
\wire{$0$} &\sn0&\xn1&\sm0&\sn0&\sm0&\sn0&\sm0&\sn0&\cxn{red}1&\sm0&\sn0&\sm0&\sn0&\sm0& \wire{$g[8,10]$} &\sm0&\cxn{darkgreen}1&\sm0&\sn0&\sn0&\sm0&\sn0&\sm0&\sn0&\sm0&\sn0&\sn0& \wire{$s_{10}$} \\
\end{tabular}

\end{center}
\caption{Out-of-place \QCLA\ adder for 10 bits.  $P$-rounds and $P^{-1}$-rounds are shown in blue.  $G$-rounds are red, and $C$-rounds are green.}
\label{oop-fig}
\end{figure}

The key relation is that the sum $s$ is equal to
$a \xor b \xor c$, where $c$ is
the carry string.  Hence, the key step in our algorithm is to compute $c$,
using the technique of the previous section.  We compute the carry string
$c_1$ through $c_n$ into the bits $Z[1]$ through $Z[n]$.

The out-of-place \QCLA\ adder proceeds as follows:
\begin{enumerate}
\item For $0 \le i < n$, $Z[i+1] \xoreq A[i]B[i]$.  This sets $z_{i+1} = g[i,i+1]$.
\item For $1 \le i < n$, $B[i] \xoreq A[i]$.
This sets $B[i] = p[i,i+1]$ for $i > 0$, which is what we need to run our
addition circuit.
\item \label{main-oop-step}
Run the circuit of Section~\ref{reversible-carry-status-sec}, using
$X$ as ancillary space.  Upon completion, $Z[i] = c_i$ for $i \ge 1$.
\item For $0 \le i < n$, $Z[i] \xoreq B[i]$.
Now, for $i > 0$, $Z[i] = a_i \xor b_i \xor c_i = s_i$.  For $i = 0$,
we have $Z[i] = b_i$.
\item Set $Z[0] \xoreq A[0]$.  For $1 \le i < n$, $B[i] \xoreq A[i]$.
This fixes $Z[0]$, and resets $B$ to its initial value.
\end{enumerate}

Aside from Step~\ref{main-oop-step}, each step occurs in a single time
slice.  So, by~(\ref{oneway-depth}), the overall depth of the circuit is
$$
\floor{\log n} + \floor{\log{n \over 3}} + 7,
$$
where three of the time-slices contain \CNOT{s} and the rest contain
Toffolis.  For $n \le 3$, the depth is slightly lower.

By~(\ref{oneway-size}), the circuit contains
$$
5n - 3w(n) - 3 \floor{\log n} - 1
$$
Toffoli gates and $3n - 1$ controlled-NOTs.

The circuit for $n = 10$ is depicted in Figure~\ref{oop-fig}.

\subsection{Addition in place}
\label{ip-qcla-sec}

For the in-place circuit, we begin the same way as above:  we
compute the carry string $c$ into $n-1$ ancillary bits (plus one output
bit for the high bit).  The total ancillary space required is
$2n - w(n) - \floor{\log n} - 1$.  We then write the low $n$ bits of the sum on
top of $b$.  The key new step is the erasure of the low $n-1$ bits of the
carry string $c$.

\begin{figure}[p]
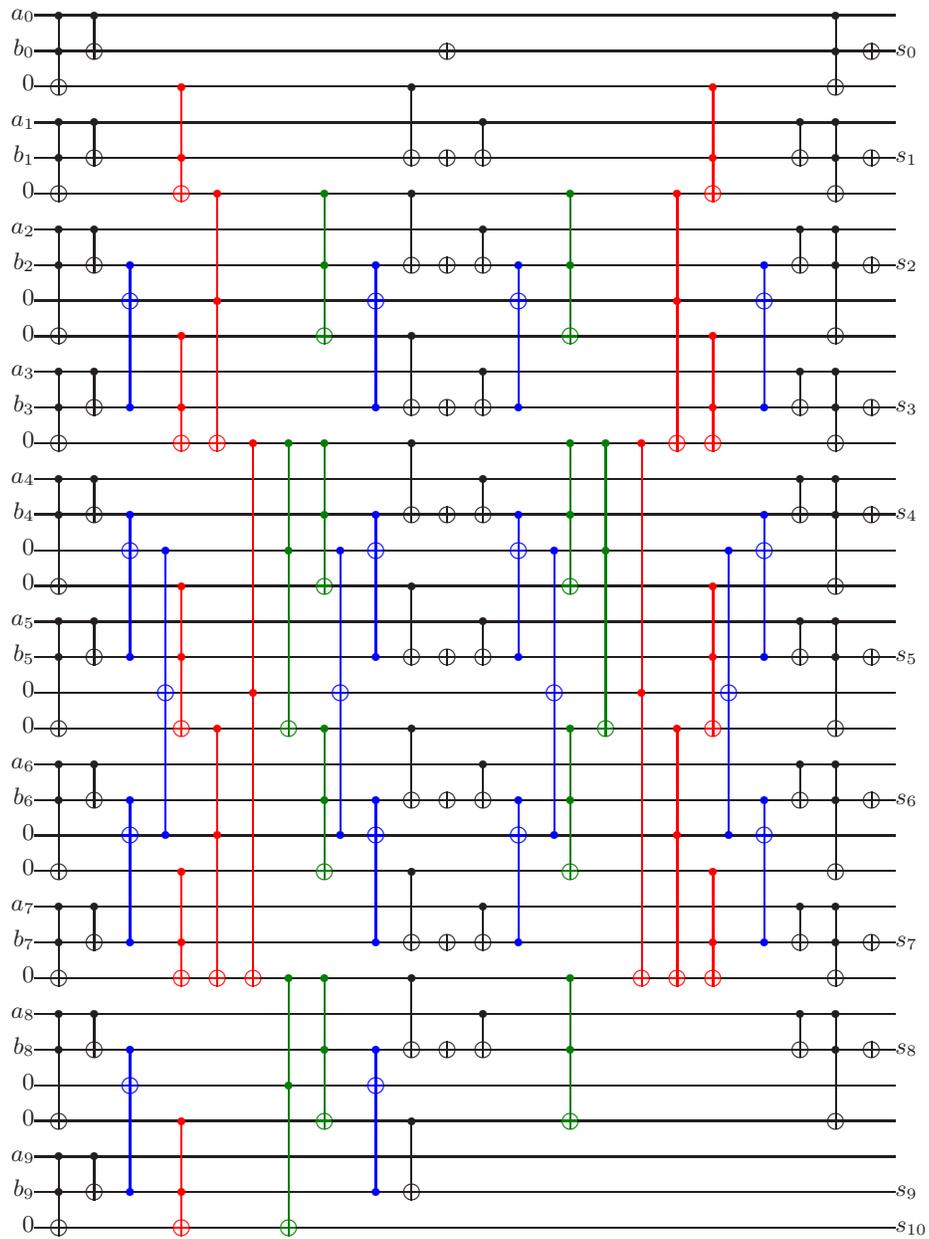

\renewcommand{\arraystretch}{0}
\renewcommand{\smwidth}{5}
\renewcommand{\smhalf}{2}
\renewcommand{\qheight}{9}
\renewcommand{\qtopheight}{4}
\renewcommand{\qtopheightminusthreehalves}{2.5}
\renewcommand{\qtopheightplustwo}{6}
\renewcommand{\qbottomheight}{5}
\renewcommand{\qbottomheightminusthreehalves}{3.5}
\renewcommand{\qbottomheightplustwo}{7}
\begin{tabular}{r@{}*{49}{c@{}}l}
\wire{$a_0$} &\sn0&\cn2&\sm0&\cn2&\sm0&\sn0&\sm0&\sn0&\sn0&\sm0&\sn0&\sm0&\sn0&\sm0&\sn0&\sm0&\sn0&\sn0&\sm0&\sn0&\sm0&\sn0&\sm0&\sn0&\sm0&\sn0&\sm0&\sn0&\sm0&\sn0&\sn0&\sm0&\sn0&\sm0&\sn0&\sm0&\sn0&\sm0&\sn0&\sn0&\sm0&\sn0&\sm0&\sn0&\sm0&\cn2&\sm0&\sn0&\sn0& \\
\wire{$b_0$} &\sn0&\cn0&\sm0&\xn1&\sm0&\sn0&\sm0&\sn0&\sn0&\sm0&\sn0&\sm0&\sn0&\sm0&\sn0&\sm0&\sn0&\sn0&\sm0&\sn0&\sm0&\sn0&\sm0&\xn3&\sm0&\sn0&\sm0&\sn0&\sm0&\sn0&\sn0&\sm0&\sn0&\sm0&\sn0&\sm0&\sn0&\sm0&\sn0&\sn0&\sm0&\sn0&\sm0&\sn0&\sm0&\cn0&\sm0&\xn3&\sn0& \wire{$s_0$} \\
\wire{$0$} &\sn0&\xn1&\sm0&\sn0&\sm0&\sn0&\sm0&\sn0&\ccn{red}2&\sm0&\sn0&\sm0&\sn0&\sm0&\sn0&\sm0&\sn0&\sn0&\sm0&\sn0&\sm0&\cn2&\sm0&\sn0&\sm0&\sn0&\sm0&\sn0&\sm0&\sn0&\sn0&\sm0&\sn0&\sm0&\sn0&\sm0&\sn0&\sm0&\ccn{red}2&\sn0&\sm0&\sn0&\sm0&\sn0&\sm0&\xn1&\sm0&\sn0&\sn0& \\
\wire{$a_1$} &\sn0&\cn2&\sm0&\cn2&\sm0&\sn0&\sm0&\sn0&\csn{red}2&\sm0&\sn0&\sm0&\sn0&\sm0&\sn0&\sm0&\sn0&\sn0&\sm0&\sn0&\sm0&\sn2&\sm0&\sn0&\sm0&\cn2&\sm0&\sn0&\sm0&\sn0&\sn0&\sm0&\sn0&\sm0&\sn0&\sm0&\sn0&\sm0&\csn{red}2&\sn0&\sm0&\sn0&\sm0&\cn2&\sm0&\cn2&\sm0&\sn0&\sn0& \\
\wire{$b_1$} &\sn0&\cn0&\sm0&\xn1&\sm0&\sn0&\sm0&\sn0&\ccn{red}0&\sm0&\sn0&\sm0&\sn0&\sm0&\sn0&\sm0&\sn0&\sn0&\sm0&\sn0&\sm0&\xn1&\sm0&\xn3&\sm0&\xn1&\sm0&\sn0&\sm0&\sn0&\sn0&\sm0&\sn0&\sm0&\sn0&\sm0&\sn0&\sm0&\ccn{red}0&\sn0&\sm0&\sn0&\sm0&\xn1&\sm0&\cn0&\sm0&\xn3&\sn0& \wire{$s_1$} \\
\wire{$0$} &\sn0&\xn1&\sm0&\sn0&\sm0&\sn0&\sm0&\sn0&\cxn{red}1&\sm0&\ccn{red}2&\sm0&\sn0&\sm0&\sn0&\sm0&\ccn{darkgreen}2&\sn0&\sm0&\sn0&\sm0&\cn2&\sm0&\sn0&\sm0&\sn0&\sm0&\sn0&\sm0&\sn0&\ccn{darkgreen}2&\sm0&\sn0&\sm0&\sn0&\sm0&\ccn{red}2&\sm0&\cxn{red}1&\sn0&\sm0&\sn0&\sm0&\sn0&\sm0&\xn1&\sm0&\sn0&\sn0& \\
\wire{$a_2$} &\sn0&\cn2&\sm0&\cn2&\sm0&\sn0&\sm0&\sn0&\sn0&\sm0&\csn{red}2&\sm0&\sn0&\sm0&\sn0&\sm0&\csn{darkgreen}2&\sn0&\sm0&\sn0&\sm0&\sn2&\sm0&\sn0&\sm0&\cn2&\sm0&\sn0&\sm0&\sn0&\csn{darkgreen}2&\sm0&\sn0&\sm0&\sn0&\sm0&\csn{red}2&\sm0&\sn0&\sn0&\sm0&\sn0&\sm0&\cn2&\sm0&\cn2&\sm0&\sn0&\sn0& \\
\wire{$b_2$} &\sn0&\cn0&\sm0&\xn1&\sm0&\ccn{blue}2&\sm0&\sn0&\sn0&\sm0&\csn{red}2&\sm0&\sn0&\sm0&\sn0&\sm0&\ccn{darkgreen}0&\sn0&\sm0&\ccn{blue}2&\sm0&\xn1&\sm0&\xn3&\sm0&\xn1&\sm0&\ccn{blue}2&\sm0&\sn0&\ccn{darkgreen}0&\sm0&\sn0&\sm0&\sn0&\sm0&\csn{red}2&\sm0&\sn0&\sn0&\sm0&\ccn{blue}2&\sm0&\xn1&\sm0&\cn0&\sm0&\xn3&\sn0& \wire{$s_2$} \\
\wire{$0$} &\sn0&\sn2&\sm0&\sn0&\sm0&\cxn{blue}0&\sm0&\sn0&\sn0&\sm0&\ccn{red}0&\sm0&\sn0&\sm0&\sn0&\sm0&\csn{darkgreen}2&\sn0&\sm0&\cxn{blue}0&\sm0&\sn0&\sm0&\sn0&\sm0&\sn0&\sm0&\cxn{blue}0&\sm0&\sn0&\csn{darkgreen}2&\sm0&\sn0&\sm0&\sn0&\sm0&\ccn{red}0&\sm0&\sn0&\sn0&\sm0&\cxn{blue}0&\sm0&\sn0&\sm0&\sn2&\sm0&\sn0&\sn0& \\
\wire{$0$} &\sn0&\xn1&\sm0&\sn0&\sm0&\csn{blue}2&\sm0&\sn0&\ccn{red}2&\sm0&\csn{red}2&\sm0&\sn0&\sm0&\sn0&\sm0&\cxn{darkgreen}1&\sn0&\sm0&\csn{blue}2&\sm0&\cn2&\sm0&\sn0&\sm0&\sn0&\sm0&\csn{blue}2&\sm0&\sn0&\cxn{darkgreen}1&\sm0&\sn0&\sm0&\sn0&\sm0&\csn{red}2&\sm0&\ccn{red}2&\sn0&\sm0&\csn{blue}2&\sm0&\sn0&\sm0&\xn1&\sm0&\sn0&\sn0& \\
\wire{$a_3$} &\sn0&\cn2&\sm0&\cn2&\sm0&\csn{blue}2&\sm0&\sn0&\csn{red}2&\sm0&\csn{red}2&\sm0&\sn0&\sm0&\sn0&\sm0&\sn0&\sn0&\sm0&\csn{blue}2&\sm0&\sn2&\sm0&\sn0&\sm0&\cn2&\sm0&\csn{blue}2&\sm0&\sn0&\sn0&\sm0&\sn0&\sm0&\sn0&\sm0&\csn{red}2&\sm0&\csn{red}2&\sn0&\sm0&\csn{blue}2&\sm0&\cn2&\sm0&\cn2&\sm0&\sn0&\sn0& \\
\wire{$b_3$} &\sn0&\cn0&\sm0&\xn1&\sm0&\ccn{blue}1&\sm0&\sn0&\ccn{red}0&\sm0&\csn{red}2&\sm0&\sn0&\sm0&\sn0&\sm0&\sn0&\sn0&\sm0&\ccn{blue}1&\sm0&\xn1&\sm0&\xn3&\sm0&\xn1&\sm0&\ccn{blue}1&\sm0&\sn0&\sn0&\sm0&\sn0&\sm0&\sn0&\sm0&\csn{red}2&\sm0&\ccn{red}0&\sn0&\sm0&\ccn{blue}1&\sm0&\xn1&\sm0&\cn0&\sm0&\xn3&\sn0& \wire{$s_3$} \\
\wire{$0$} &\sn0&\xn1&\sm0&\sn0&\sm0&\sn0&\sm0&\sn0&\cxn{red}1&\sm0&\cxn{red}1&\sm0&\ccn{red}2&\sm0&\ccn{darkgreen}2&\sm0&\ccn{darkgreen}2&\sn0&\sm0&\sn0&\sm0&\cn2&\sm0&\sn0&\sm0&\sn0&\sm0&\sn0&\sm0&\sn0&\ccn{darkgreen}2&\sm0&\ccn{darkgreen}2&\sm0&\ccn{red}2&\sm0&\cxn{red}1&\sm0&\cxn{red}1&\sn0&\sm0&\sn0&\sm0&\sn0&\sm0&\xn1&\sm0&\sn0&\sn0& \\
\wire{$a_4$} &\sn0&\cn2&\sm0&\cn2&\sm0&\sn0&\sm0&\sn0&\sn0&\sm0&\sn0&\sm0&\csn{red}2&\sm0&\csn{darkgreen}2&\sm0&\csn{darkgreen}2&\sn0&\sm0&\sn0&\sm0&\sn2&\sm0&\sn0&\sm0&\cn2&\sm0&\sn0&\sm0&\sn0&\csn{darkgreen}2&\sm0&\csn{darkgreen}2&\sm0&\csn{red}2&\sm0&\sn0&\sm0&\sn0&\sn0&\sm0&\sn0&\sm0&\cn2&\sm0&\cn2&\sm0&\sn0&\sn0& \\
\wire{$b_4$} &\sn0&\cn0&\sm0&\xn1&\sm0&\ccn{blue}2&\sm0&\sn0&\sn0&\sm0&\sn0&\sm0&\csn{red}2&\sm0&\csn{darkgreen}2&\sm0&\ccn{darkgreen}0&\sn0&\sm0&\ccn{blue}2&\sm0&\xn1&\sm0&\xn3&\sm0&\xn1&\sm0&\ccn{blue}2&\sm0&\sn0&\ccn{darkgreen}0&\sm0&\csn{darkgreen}2&\sm0&\csn{red}2&\sm0&\sn0&\sm0&\sn0&\sn0&\sm0&\ccn{blue}2&\sm0&\xn1&\sm0&\cn0&\sm0&\xn3&\sn0& \wire{$s_4$} \\
\wire{$0$} &\sn0&\sn2&\sm0&\sn0&\sm0&\cxn{blue}0&\sm0&\ccn{blue}2&\sn0&\sm0&\sn0&\sm0&\csn{red}2&\sm0&\ccn{darkgreen}0&\sm0&\csn{darkgreen}2&\ccn{blue}2&\sm0&\cxn{blue}0&\sm0&\sn0&\sm0&\sn0&\sm0&\sn0&\sm0&\cxn{blue}0&\sm0&\ccn{blue}2&\csn{darkgreen}2&\sm0&\ccn{darkgreen}0&\sm0&\csn{red}2&\sm0&\sn0&\sm0&\sn0&\ccn{blue}2&\sm0&\cxn{blue}0&\sm0&\sn0&\sm0&\sn2&\sm0&\sn0&\sn0& \\
\wire{$0$} &\sn0&\xn1&\sm0&\sn0&\sm0&\csn{blue}2&\sm0&\csn{blue}2&\ccn{red}2&\sm0&\sn0&\sm0&\csn{red}2&\sm0&\csn{darkgreen}2&\sm0&\cxn{darkgreen}1&\csn{blue}2&\sm0&\csn{blue}2&\sm0&\cn2&\sm0&\sn0&\sm0&\sn0&\sm0&\csn{blue}2&\sm0&\csn{blue}2&\cxn{darkgreen}1&\sm0&\csn{darkgreen}2&\sm0&\csn{red}2&\sm0&\sn0&\sm0&\ccn{red}2&\csn{blue}2&\sm0&\csn{blue}2&\sm0&\sn0&\sm0&\xn1&\sm0&\sn0&\sn0& \\
\wire{$a_5$} &\sn0&\cn2&\sm0&\cn2&\sm0&\csn{blue}2&\sm0&\csn{blue}2&\csn{red}2&\sm0&\sn0&\sm0&\csn{red}2&\sm0&\csn{darkgreen}2&\sm0&\sn0&\csn{blue}2&\sm0&\csn{blue}2&\sm0&\sn2&\sm0&\sn0&\sm0&\cn2&\sm0&\csn{blue}2&\sm0&\csn{blue}2&\sn0&\sm0&\csn{darkgreen}2&\sm0&\csn{red}2&\sm0&\sn0&\sm0&\csn{red}2&\csn{blue}2&\sm0&\csn{blue}2&\sm0&\cn2&\sm0&\cn2&\sm0&\sn0&\sn0& \\
\wire{$b_5$} &\sn0&\cn0&\sm0&\xn1&\sm0&\ccn{blue}1&\sm0&\csn{blue}2&\ccn{red}0&\sm0&\sn0&\sm0&\csn{red}2&\sm0&\csn{darkgreen}2&\sm0&\sn0&\csn{blue}2&\sm0&\ccn{blue}1&\sm0&\xn1&\sm0&\xn3&\sm0&\xn1&\sm0&\ccn{blue}1&\sm0&\csn{blue}2&\sn0&\sm0&\csn{darkgreen}2&\sm0&\csn{red}2&\sm0&\sn0&\sm0&\ccn{red}0&\csn{blue}2&\sm0&\ccn{blue}1&\sm0&\xn1&\sm0&\cn0&\sm0&\xn3&\sn0& \wire{$s_5$} \\
\wire{$0$} &\sn0&\sn2&\sm0&\sn0&\sm0&\sn0&\sm0&\cxn{blue}0&\csn{red}2&\sm0&\sn0&\sm0&\ccn{red}0&\sm0&\csn{darkgreen}2&\sm0&\sn0&\cxn{blue}0&\sm0&\sn0&\sm0&\sn0&\sm0&\sn0&\sm0&\sn0&\sm0&\sn0&\sm0&\cxn{blue}0&\sn0&\sm0&\csn{darkgreen}2&\sm0&\ccn{red}0&\sm0&\sn0&\sm0&\csn{red}2&\cxn{blue}0&\sm0&\sn0&\sm0&\sn0&\sm0&\sn2&\sm0&\sn0&\sn0& \\
\wire{$0$} &\sn0&\xn1&\sm0&\sn0&\sm0&\sn0&\sm0&\csn{blue}2&\cxn{red}1&\sm0&\ccn{red}2&\sm0&\csn{red}2&\sm0&\cxn{darkgreen}1&\sm0&\ccn{darkgreen}2&\csn{blue}2&\sm0&\sn0&\sm0&\cn2&\sm0&\sn0&\sm0&\sn0&\sm0&\sn0&\sm0&\csn{blue}2&\ccn{darkgreen}2&\sm0&\cxn{darkgreen}1&\sm0&\csn{red}2&\sm0&\ccn{red}2&\sm0&\cxn{red}1&\csn{blue}2&\sm0&\sn0&\sm0&\sn0&\sm0&\xn1&\sm0&\sn0&\sn0& \\
\wire{$a_6$} &\sn0&\cn2&\sm0&\cn2&\sm0&\sn0&\sm0&\csn{blue}2&\sn0&\sm0&\csn{red}2&\sm0&\csn{red}2&\sm0&\sn0&\sm0&\csn{darkgreen}2&\csn{blue}2&\sm0&\sn0&\sm0&\sn2&\sm0&\sn0&\sm0&\cn2&\sm0&\sn0&\sm0&\csn{blue}2&\csn{darkgreen}2&\sm0&\sn0&\sm0&\csn{red}2&\sm0&\csn{red}2&\sm0&\sn0&\csn{blue}2&\sm0&\sn0&\sm0&\cn2&\sm0&\cn2&\sm0&\sn0&\sn0& \\
\wire{$b_6$} &\sn0&\cn0&\sm0&\xn1&\sm0&\ccn{blue}2&\sm0&\csn{blue}2&\sn0&\sm0&\csn{red}2&\sm0&\csn{red}2&\sm0&\sn0&\sm0&\ccn{darkgreen}0&\csn{blue}2&\sm0&\ccn{blue}2&\sm0&\xn1&\sm0&\xn3&\sm0&\xn1&\sm0&\ccn{blue}2&\sm0&\csn{blue}2&\ccn{darkgreen}0&\sm0&\sn0&\sm0&\csn{red}2&\sm0&\csn{red}2&\sm0&\sn0&\csn{blue}2&\sm0&\ccn{blue}2&\sm0&\xn1&\sm0&\cn0&\sm0&\xn3&\sn0& \wire{$s_6$} \\
\wire{$0$} &\sn0&\sn2&\sm0&\sn0&\sm0&\cxn{blue}0&\sm0&\ccn{blue}1&\sn0&\sm0&\ccn{red}0&\sm0&\csn{red}2&\sm0&\sn0&\sm0&\csn{darkgreen}2&\ccn{blue}1&\sm0&\cxn{blue}0&\sm0&\sn0&\sm0&\sn0&\sm0&\sn0&\sm0&\cxn{blue}0&\sm0&\ccn{blue}1&\csn{darkgreen}2&\sm0&\sn0&\sm0&\csn{red}2&\sm0&\ccn{red}0&\sm0&\sn0&\ccn{blue}1&\sm0&\cxn{blue}0&\sm0&\sn0&\sm0&\sn2&\sm0&\sn0&\sn0& \\
\wire{$0$} &\sn0&\xn1&\sm0&\sn0&\sm0&\csn{blue}2&\sm0&\sn0&\ccn{red}2&\sm0&\csn{red}2&\sm0&\csn{red}2&\sm0&\sn0&\sm0&\cxn{darkgreen}1&\sn0&\sm0&\csn{blue}2&\sm0&\cn2&\sm0&\sn0&\sm0&\sn0&\sm0&\csn{blue}2&\sm0&\sn0&\cxn{darkgreen}1&\sm0&\sn0&\sm0&\csn{red}2&\sm0&\csn{red}2&\sm0&\ccn{red}2&\sn0&\sm0&\csn{blue}2&\sm0&\sn0&\sm0&\xn1&\sm0&\sn0&\sn0& \\
\wire{$a_7$} &\sn0&\cn2&\sm0&\cn2&\sm0&\csn{blue}2&\sm0&\sn0&\csn{red}2&\sm0&\csn{red}2&\sm0&\csn{red}2&\sm0&\sn0&\sm0&\sn0&\sn0&\sm0&\csn{blue}2&\sm0&\sn2&\sm0&\sn0&\sm0&\cn2&\sm0&\csn{blue}2&\sm0&\sn0&\sn0&\sm0&\sn0&\sm0&\csn{red}2&\sm0&\csn{red}2&\sm0&\csn{red}2&\sn0&\sm0&\csn{blue}2&\sm0&\cn2&\sm0&\cn2&\sm0&\sn0&\sn0& \\
\wire{$b_7$} &\sn0&\cn0&\sm0&\xn1&\sm0&\ccn{blue}1&\sm0&\sn0&\ccn{red}0&\sm0&\csn{red}2&\sm0&\csn{red}2&\sm0&\sn0&\sm0&\sn0&\sn0&\sm0&\ccn{blue}1&\sm0&\xn1&\sm0&\xn3&\sm0&\xn1&\sm0&\ccn{blue}1&\sm0&\sn0&\sn0&\sm0&\sn0&\sm0&\csn{red}2&\sm0&\csn{red}2&\sm0&\ccn{red}0&\sn0&\sm0&\ccn{blue}1&\sm0&\xn1&\sm0&\cn0&\sm0&\xn3&\sn0& \wire{$s_7$} \\
\wire{$0$} &\sn0&\xn1&\sm0&\sn0&\sm0&\sn0&\sm0&\sn0&\cxn{red}1&\sm0&\cxn{red}1&\sm0&\cxn{red}1&\sm0&\ccn{darkgreen}2&\sm0&\ccn{darkgreen}2&\sn0&\sm0&\sn0&\sm0&\cn2&\sm0&\sn0&\sm0&\sn0&\sm0&\sn0&\sm0&\sn0&\ccn{darkgreen}2&\sm0&\sn0&\sm0&\cxn{red}1&\sm0&\cxn{red}1&\sm0&\cxn{red}1&\sn0&\sm0&\sn0&\sm0&\sn0&\sm0&\xn1&\sm0&\sn0&\sn0& \\
\wire{$a_8$} &\sn0&\cn2&\sm0&\cn2&\sm0&\sn0&\sm0&\sn0&\sn0&\sm0&\sn0&\sm0&\sn0&\sm0&\csn{darkgreen}2&\sm0&\csn{darkgreen}2&\sn0&\sm0&\sn0&\sm0&\sn2&\sm0&\sn0&\sm0&\cn2&\sm0&\sn0&\sm0&\sn0&\csn{darkgreen}2&\sm0&\sn0&\sm0&\sn0&\sm0&\sn0&\sm0&\sn0&\sn0&\sm0&\sn0&\sm0&\cn2&\sm0&\cn2&\sm0&\sn0&\sn0& \\
\wire{$b_8$} &\sn0&\cn0&\sm0&\xn1&\sm0&\ccn{blue}2&\sm0&\sn0&\sn0&\sm0&\sn0&\sm0&\sn0&\sm0&\csn{darkgreen}2&\sm0&\ccn{darkgreen}0&\sn0&\sm0&\ccn{blue}2&\sm0&\xn1&\sm0&\xn3&\sm0&\xn1&\sm0&\sn0&\sm0&\sn0&\ccn{darkgreen}0&\sm0&\sn0&\sm0&\sn0&\sm0&\sn0&\sm0&\sn0&\sn0&\sm0&\sn0&\sm0&\xn1&\sm0&\cn0&\sm0&\xn3&\sn0& \wire{$s_8$} \\
\wire{$0$} &\sn0&\sn2&\sm0&\sn0&\sm0&\cxn{blue}0&\sm0&\sn0&\sn0&\sm0&\sn0&\sm0&\sn0&\sm0&\ccn{darkgreen}0&\sm0&\csn{darkgreen}2&\sn0&\sm0&\cxn{blue}0&\sm0&\sn0&\sm0&\sn0&\sm0&\sn0&\sm0&\sn0&\sm0&\sn0&\csn{darkgreen}2&\sm0&\sn0&\sm0&\sn0&\sm0&\sn0&\sm0&\sn0&\sn0&\sm0&\sn0&\sm0&\sn0&\sm0&\sn2&\sm0&\sn0&\sn0& \\
\wire{$0$} &\sn0&\xn1&\sm0&\sn0&\sm0&\csn{blue}2&\sm0&\sn0&\ccn{red}2&\sm0&\sn0&\sm0&\sn0&\sm0&\csn{darkgreen}2&\sm0&\cxn{darkgreen}1&\sn0&\sm0&\csn{blue}2&\sm0&\cn2&\sm0&\sn0&\sm0&\sn0&\sm0&\sn0&\sm0&\sn0&\cxn{darkgreen}1&\sm0&\sn0&\sm0&\sn0&\sm0&\sn0&\sm0&\sn0&\sn0&\sm0&\sn0&\sm0&\sn0&\sm0&\xn1&\sm0&\sn0&\sn0& \\
\wire{$a_9$} &\sn0&\cn2&\sm0&\cn2&\sm0&\csn{blue}2&\sm0&\sn0&\csn{red}2&\sm0&\sn0&\sm0&\sn0&\sm0&\csn{darkgreen}2&\sm0&\sn0&\sn0&\sm0&\csn{blue}2&\sm0&\sn2&\sm0&\sn0&\sm0&\sn0&\sm0&\sn0&\sm0&\sn0&\sn0&\sm0&\sn0&\sm0&\sn0&\sm0&\sn0&\sm0&\sn0&\sn0&\sm0&\sn0&\sm0&\sn0&\sm0&\sn0&\sm0&\sn0&\sn0& \\
\wire{$b_9$} &\sn0&\cn0&\sm0&\xn1&\sm0&\ccn{blue}1&\sm0&\sn0&\ccn{red}0&\sm0&\sn0&\sm0&\sn0&\sm0&\csn{darkgreen}2&\sm0&\sn0&\sn0&\sm0&\ccn{blue}1&\sm0&\xn1&\sm0&\sn0&\sm0&\sn0&\sm0&\sn0&\sm0&\sn0&\sn0&\sm0&\sn0&\sm0&\sn0&\sm0&\sn0&\sm0&\sn0&\sn0&\sm0&\sn0&\sm0&\sn0&\sm0&\sn0&\sm0&\sn0&\sn0& \wire{$s_9$} \\
\wire{$0$} &\sn0&\xn1&\sm0&\sn0&\sm0&\sn0&\sm0&\sn0&\cxn{red}1&\sm0&\sn0&\sm0&\sn0&\sm0&\cxn{darkgreen}1&\sm0&\sn0&\sn0&\sm0&\sn0&\sm0&\sn0&\sm0&\sn0&\sm0&\sn0&\sm0&\sn0&\sm0&\sn0&\sn0&\sm0&\sn0&\sm0&\sn0&\sm0&\sn0&\sm0&\sn0&\sn0&\sm0&\sn0&\sm0&\sn0&\sm0&\sn0&\sm0&\sn0&\sn0& \wire{$s_{10}$} \\
\end{tabular}

\caption{In-place \QCLA\ adder for 10 bits. $P$-rounds and
$P^{-1}$-rounds are shown in blue.  $G$-rounds are red,
and $C$-rounds are green.}
\label{ip-fig}
\end{figure}

Recall from Section~\ref{notation-sec} that we are using
two's-complement arithmetic:
$$
r' + r \equiv -1 \pmod{2^n}.
$$
So, writing $s = a+b$,
$$
a + s' \equiv a - a - b - 1 \equiv b' \pmod{2^n}.
$$
Let $d$ be the carry string generated by $a$ and $s'$.  We have
\begin{align*}
a \xor s' \xor d & = b'\\
a \xor (a \xor b \xor c) \xor (-1) \xor d & = b \xor (-1)\\
c &= d.
\end{align*}
So the carry string $d$, generated by adding $a$ and $s'$, is simply $c$.
After we compute $s$, we can complement it, and then run the circuit
of Section~\ref{reversible-carry-status-sec} in reverse to erase $c$.

The in-place \QCLA\ adder proceeds as follows.  We denote
the $n-1$ ancillae which store the carry string as $Z[1], \dots, Z[n-1]$,
and the remaining ancillae as $X$.  The output bit is labeled $Z[n]$.

\begin{enumerate}
\item\label{first-ip-step} For $0 \le i < n$, $Z[i+1] \xoreq A[i] B[i]$.
This sets $Z[i+1] = g[i,i+1]$.
\item For $0 \le i < n$, $B[i] \xoreq A[i]$.
This sets $B[i] = p[i,i+1]$ for $i > 0$.  Also, $B[0] = s_0$.
\item \label{main-ip-step-1}
Run the circuit of Section~\ref{reversible-carry-status-sec}, using
$X$ as ancillary space.  Upon completion, $Z[i] = c_i$ for $i \ge 1$.
\item For $1 \le i < n$, $B[i] \xoreq Z[i]$.  Now $B[i] = s_i$.
\item For $0 \le i < n-1$, negate $B[i]$.  Now $B$ contains $s'$.
\item For $1 \le i < n-1$, $B[i] \xoreq A[i]$.
\item \label{main-ip-step-2} Run the circuit of Section~\ref{reversible-carry-status-sec} in reverse.\footnote{In Step~\ref{main-ip-step-2}, we actually
reverse the $(n-1)$-bit adder, since we should not erase the high carry
bit. See Section~\ref{mod-2n-sec} for more discussion.}
Upon completion, $Z[i+1] = a_i s'_i$ for $0 \le i < n-1$,
and $B[i] = a_i \xor s'_i$ for $1 \le i < n$.
\item For $1 \le i < n-1$, $B[i] \xoreq A[i]$.
\item For $0 \le i < n-1$, $Z[i+1] \xoreq A[i] B[i]$.
\item For $0 \le i < n-1$, negate $B[i]$.
\end{enumerate}

Each step other than~\ref{main-ip-step-1} and~\ref{main-ip-step-2}
has depth 1.  By~(\ref{oneway-depth}), the overall depth is
$$
\floor{\log n} + \floor{\log (n-1)} + \floor{\log{n\over3}}
+ \floor{\log{n-1\over3}} + 14,
$$
where two of the time-slices contain negations, four contain
\CNOT{s}, and the rest contain Toffolis.  For some values of
$n \le 6$, the depth is slightly lower.

By~(\ref{oneway-size}), the circuit contains
$$
10n - 3 w(n) - 3 w(n-1) - 3 \floor{\log n} - 3 \floor{\log(n-1)} - 7
$$
Toffoli gates, $4n - 5$ controlled-NOTs, and $2n - 2$ negations.

In Figure \ref{ip-fig},
we show a sample in-place \QCLA\ adder for the case $n=10$. 

\section{Extensions}
\label{extensions-sec}

We now discuss various modified versions of the circuit.  The simplest
is one which adds (mod $2^n$); we simply skip the computation of the
high bit.  With slightly more work, we can add (mod $2^n - 1$).

There are other constructions which use the log-depth
adder as a subroutine.  For these, it is useful
to allow the adder to take one additional bit, an incoming carry.
In this case, we wish to compute $a + b + y$, where $y$ is either 0 or 1.

Another useful subroutine in an addition (or modular addition)
circuit is comparison:  Is $a \ge b$?  Equivalently, is the
high bit of $a - b$ zero?  We discuss how one can use the log-depth
adder to subtract, and we show that a comparator is of comparable
complexity to an out-of-place adder.

\subsection{Addition \protect\boldmath{(mod $2^n$)}}
\label{mod-2n-sec}

It is straightforward to add (mod $2^n$); we simply do not compute the high
bit of the sum.  The only question is:  what are the exact savings,
in depth and circuit size?

Since we do not need to compute $c_n$, we can simply run the circuit of
Section~\ref{reversible-carry-status-sec} on the low-order $n-1$ bits
of $a$ and $b$.  For the out-of-place adder, this circuit
leaves $c_{n-1}$ in $Z[n-1]$, so we also need to apply the
gates $Z[n-1] \xoreq a_{n-1}$
and $Z[n-1] \xoreq b_{n-1}$; we therefore add two additional \CNOT{s}.
For $n > 1$, this does not increase the depth.

So, the out-of-place (mod $2^n$) adder produces $n$ output bits,
and uses $(n-1) - w(n-1) - \floor{\log (n-1)}$ ancillae.
The depth is $\floor{\log(n-1)} + \floor{\log{n - 1 \over 3}} + 7$
when $n \ge 4$, and the circuit consists of $5n - 3w(n-1)
- 3 \floor{\log (n - 1)} - 6$ Toffolis and $3n - 2$ controlled-NOTs.

For the in-place adder, we follow the steps in Section~\ref{ip-qcla-sec}.
However, in Step~\ref{first-ip-step}, our loop now stops at $i = n-2$,
and, in Step~\ref{main-ip-step-1}, we run the $(n-1)$-bit adder.

Thus, the in-place (mod $2^n$) adder uses
$2n - 2 - w(n-1) - \floor{\log (n-1)}$ ancillae.  The depth
is $2\floor{\log(n-1)} + 2 \floor{\log{n-1\over3}} + 14$ when
$n \ge 5$, and the circuit size is $10n - 6w(n-1) - 6\floor{\log(n-1)}
- 12$ Toffolis, $4n - 5$ controlled-NOTs, and $2n - 2$ negations.

\subsection{Addition with incoming carry}
\label{incoming-carry-sec}

Suppose we want our adder to take $2n + 1$ bits of input:  $a$, $b$, and
a single bit $y$, representing an incoming carry.  This is useful in
various hybrid addition circuits, where we break the problem up into
smaller pieces.  

We can accomplish this by adding the $(n+1)$-bit numbers
$2a + y$ and $2b + y$, whose sum is $2(a + b + y)$.  So, the cost is
roughly the same as that of an $(n+1)$-bit add.
However, we use fewer operations
on the low-order bit---we simply start with $c_1 = y$.  The additional
input bit $y$ replaces one output bit for the out-of-place adder, and
one ancillary bit for the in-place adder.

For the out-of-place adder, we save one Toffoli and two controlled-NOTs
over the usual $(n+1)$-bit adder.  For the in-place adder, we save
two Toffolis, one controlled-NOT, and two negations.

The same analysis applies to the (mod $2^n$) adder of Section~\ref{mod-2n-sec}.

\subsection{Subtraction}
\label{subtraction-sec}

It is straightforward to use our circuit to compute $a - b$.  First,
complement all bits of $a$.  Then, add as usual; we compute
$a' + b$.  At the end, complement all bits of $a$ and all output bits.
The result, assuming two's-complement arithmetic, is then
$$
(a' + b)' = (-a - 1 + b)' = a - b.
$$
A similar argument holds for one's-complement arithmetic.

Hence, the cost of subtraction is essentially the same as the cost
of addition.  We add two time-slices, both consisting only of
negations.

\subsection{Comparison}
\label{comparison-sec}
Suppose we wish to compare two numbers $a$ and $b$.  We
compute the high bit of $a - b$.  As in the subtractor, we first
complement $a$.  We then run the \QCLA\ adder forward until we
have found the high bit of $a' + b$, and then we reverse the
preceding computation.  This gives us a \QCLA\ comparator.

\enlargethispage{\baselineskip}
If $n = 2^k$ for some $k$, then the above idea works well; we find the high
bit at the end of the $G$-rounds, halfway through the out-of-place
add.  However, for $n = 2^k - 1$, we do not compute the high bit until
after we're done with the $C$-rounds.  If we just use this simple
approach, the depth of our circuit turns out to be $2 \floor{\log n}
+ 2w(n) + 5$.  We would prefer to design a comparator which has depth
$2 \log n + O(1)$.

So, we have to be more careful.  Let $k = \ceil{\log n}$.  If we
just do and undo the $P$-rounds and $G$-rounds, we can compare two
$2^k$-bit numbers in depth roughly $2k$.  So, we can pad our
$n$-bit numbers by adding zeros to the front, and then use the
compare circuit for $2^k$-bit numbers.

After we complement $a$, we will have $p[i,j] = 1$ for $j > i \ge n$
and $g[i,j] = 0$ for $j > i \ge n$.  We do not explicitly compute these
values in our circuit; effectively, we compile the values into the circuit.

Overall, the comparator uses $2n - \floor{\log (n - 1)} - 3$ ancillae.
For the explicit discussion, we suppose our input is stored in
the $n$-long bit arrays $A$ and $B$.  We have $n-1$ ancillary bits
denoted $Z[1], \dots, Z[n-1]$, and $n - \floor{\log (n -1)} - 2$
additional ancillae denoted by $X$.  The output bit is denoted $Z[n]$.
We proceed as follows:

\begin{enumerate}
\item For $0 \le i < n$, negate $A[i]$.
\item For $0 \le i < n$, $Z[i+1] \xoreq A[i]B[i]$.
\item For $1 \le i < n$, $B[i] \xoreq A[i]$.
\item\label{compare-P-step} Do the $P$-rounds for the $2^k$-bit adder using space $X$; write
only the values we cannot deduce at compile-time.
\item\label{compare-G-step} Do the $G$-rounds for the $2^k$-bit adder; apply only those
gates which affect $Z[n]$.
\item\label{compare-G2-step} Undo the $G$-round gates which did not write to $Z[n]$.
\item\label{compare-P2-step} Do the $P^{-1}$-rounds for the $2^k$-bit adder, erasing $X$.
\item For $1 \le i < n$, $B[i] \xoreq A[i]$.
\item\label{compare-g-erase-step} For $0 \le i < n - 1$,
$Z[i+1] \xoreq A[i]B[i]$.
\item For $0 \le i < n$, negate $A[i]$.  Also negate $Z[n]$.
\end{enumerate}

Step~\ref{compare-G-step} contains $n - 1$ Toffoli gates, in depth
$\floor{\log (n - 1)} + 1$.  Step~\ref{compare-G2-step} is equivalent
to inverting the $G$-rounds for an $(n-1)$-bit adder, and contains
$n - w(n-1) - 1$ Toffoli gates in depth $\floor{\log (n - 1)}$.

Steps~\ref{compare-P-step} and~\ref{compare-P2-step} each consist
of $n - \floor{\log (n - 1)} - 2$ gates.  In each case, the depth would
be $\floor{\log (n-1)}$, but each $P$-round after the first (and
each $P^{-1}$-round before the last) can be done in parallel with
a $G$-round.

\begin{figure}
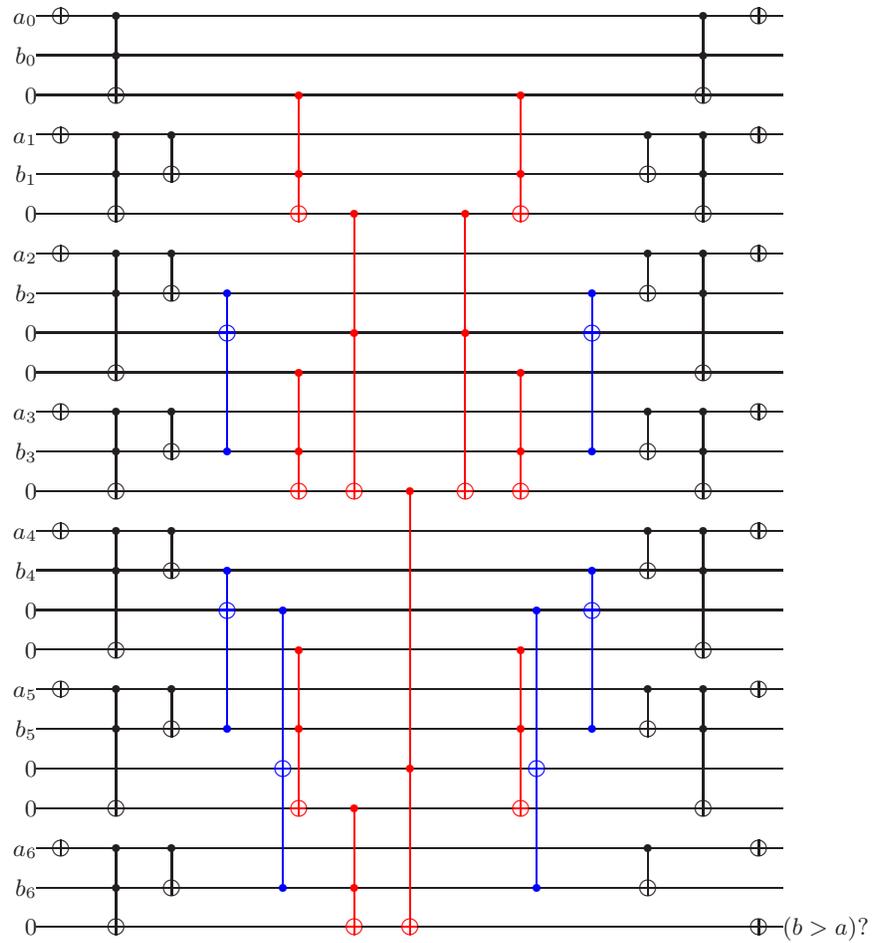

\begin{center}
\renewcommand{\arraystretch}{0}
\renewcommand{\smwidth}{10}
\renewcommand{\smhalf}{5}
\begin{tabular}{r@{}*{29}{c@{}}l}
\wire{$a_0$} &\sn0&\xn3&\sm0&\cn2&\sm0&\sn0&\sm0&\sn0&\sm0&\sn0&\sn0&\sm0&\sn0&\sm0&\sn0&\sm0&\sn0&\sm0&\sn0&\sn0&\sm0&\sn0&\sm0&\sn0&\sm0&\cn2&\sm0&\xn3&\sn0& \\
\wire{$b_0$} &\sn0&\sn0&\sm0&\cn0&\sm0&\sn0&\sm0&\sn0&\sm0&\sn0&\sn0&\sm0&\sn0&\sm0&\sn0&\sm0&\sn0&\sm0&\sn0&\sn0&\sm0&\sn0&\sm0&\sn0&\sm0&\cn0&\sm0&\sn0&\sn0& \\
\wire{$0$} &\sn0&\sn0&\sm0&\xn1&\sm0&\sn0&\sm0&\sn0&\sm0&\sn0&\ccn{red}2&\sm0&\sn0&\sm0&\sn0&\sm0&\sn0&\sm0&\ccn{red}2&\sn0&\sm0&\sn0&\sm0&\sn0&\sm0&\xn1&\sm0&\sn0&\sn0& \\
\wire{$a_1$} &\sn0&\xn3&\sm0&\cn2&\sm0&\cn2&\sm0&\sn0&\sm0&\sn0&\csn{red}2&\sm0&\sn0&\sm0&\sn0&\sm0&\sn0&\sm0&\csn{red}2&\sn0&\sm0&\sn0&\sm0&\cn2&\sm0&\cn2&\sm0&\xn3&\sn0& \\
\wire{$b_1$} &\sn0&\sn0&\sm0&\cn0&\sm0&\xn1&\sm0&\sn0&\sm0&\sn0&\ccn{red}0&\sm0&\sn0&\sm0&\sn0&\sm0&\sn0&\sm0&\ccn{red}0&\sn0&\sm0&\sn0&\sm0&\xn1&\sm0&\cn0&\sm0&\sn0&\sn0& \\
\wire{$0$} &\sn0&\sn0&\sm0&\xn1&\sm0&\sn0&\sm0&\sn0&\sm0&\sn0&\cxn{red}1&\sm0&\ccn{red}2&\sm0&\sn0&\sm0&\ccn{red}2&\sm0&\cxn{red}1&\sn0&\sm0&\sn0&\sm0&\sn0&\sm0&\xn1&\sm0&\sn0&\sn0& \\
\wire{$a_2$} &\sn0&\xn3&\sm0&\cn2&\sm0&\cn2&\sm0&\sn0&\sm0&\sn0&\sn0&\sm0&\csn{red}2&\sm0&\sn0&\sm0&\csn{red}2&\sm0&\sn0&\sn0&\sm0&\sn0&\sm0&\cn2&\sm0&\cn2&\sm0&\xn3&\sn0& \\
\wire{$b_2$} &\sn0&\sn0&\sm0&\cn0&\sm0&\xn1&\sm0&\ccn{blue}2&\sm0&\sn0&\sn0&\sm0&\csn{red}2&\sm0&\sn0&\sm0&\csn{red}2&\sm0&\sn0&\sn0&\sm0&\ccn{blue}2&\sm0&\xn1&\sm0&\cn0&\sm0&\sn0&\sn0& \\
\wire{$0$} &\sn0&\sn0&\sm0&\sn2&\sm0&\sn0&\sm0&\cxn{blue}0&\sm0&\sn0&\sn0&\sm0&\ccn{red}0&\sm0&\sn0&\sm0&\ccn{red}0&\sm0&\sn0&\sn0&\sm0&\cxn{blue}0&\sm0&\sn0&\sm0&\sn2&\sm0&\sn0&\sn0& \\
\wire{$0$} &\sn0&\sn0&\sm0&\xn1&\sm0&\sn0&\sm0&\csn{blue}2&\sm0&\sn0&\ccn{red}2&\sm0&\csn{red}2&\sm0&\sn0&\sm0&\csn{red}2&\sm0&\ccn{red}2&\sn0&\sm0&\csn{blue}2&\sm0&\sn0&\sm0&\xn1&\sm0&\sn0&\sn0& \\
\wire{$a_3$} &\sn0&\xn3&\sm0&\cn2&\sm0&\cn2&\sm0&\csn{blue}2&\sm0&\sn0&\csn{red}2&\sm0&\csn{red}2&\sm0&\sn0&\sm0&\csn{red}2&\sm0&\csn{red}2&\sn0&\sm0&\csn{blue}2&\sm0&\cn2&\sm0&\cn2&\sm0&\xn3&\sn0& \\
\wire{$b_3$} &\sn0&\sn0&\sm0&\cn0&\sm0&\xn1&\sm0&\ccn{blue}1&\sm0&\sn0&\ccn{red}0&\sm0&\csn{red}2&\sm0&\sn0&\sm0&\csn{red}2&\sm0&\ccn{red}0&\sn0&\sm0&\ccn{blue}1&\sm0&\xn1&\sm0&\cn0&\sm0&\sn0&\sn0& \\
\wire{$0$} &\sn0&\sn0&\sm0&\xn1&\sm0&\sn0&\sm0&\sn0&\sm0&\sn0&\cxn{red}1&\sm0&\cxn{red}1&\sm0&\ccn{red}2&\sm0&\cxn{red}1&\sm0&\cxn{red}1&\sn0&\sm0&\sn0&\sm0&\sn0&\sm0&\xn1&\sm0&\sn0&\sn0& \\
\wire{$a_4$} &\sn0&\xn3&\sm0&\cn2&\sm0&\cn2&\sm0&\sn0&\sm0&\sn0&\sn0&\sm0&\sn0&\sm0&\csn{red}2&\sm0&\sn0&\sm0&\sn0&\sn0&\sm0&\sn0&\sm0&\cn2&\sm0&\cn2&\sm0&\xn3&\sn0& \\
\wire{$b_4$} &\sn0&\sn0&\sm0&\cn0&\sm0&\xn1&\sm0&\ccn{blue}2&\sm0&\sn0&\sn0&\sm0&\sn0&\sm0&\csn{red}2&\sm0&\sn0&\sm0&\sn0&\sn0&\sm0&\ccn{blue}2&\sm0&\xn1&\sm0&\cn0&\sm0&\sn0&\sn0& \\
\wire{$0$} &\sn0&\sn0&\sm0&\sn2&\sm0&\sn0&\sm0&\cxn{blue}0&\sm0&\ccn{blue}2&\sn0&\sm0&\sn0&\sm0&\csn{red}2&\sm0&\sn0&\sm0&\sn0&\ccn{blue}2&\sm0&\cxn{blue}0&\sm0&\sn0&\sm0&\sn2&\sm0&\sn0&\sn0& \\
\wire{$0$} &\sn0&\sn0&\sm0&\xn1&\sm0&\sn0&\sm0&\csn{blue}2&\sm0&\csn{blue}2&\ccn{red}2&\sm0&\sn0&\sm0&\csn{red}2&\sm0&\sn0&\sm0&\ccn{red}2&\csn{blue}2&\sm0&\csn{blue}2&\sm0&\sn0&\sm0&\xn1&\sm0&\sn0&\sn0& \\
\wire{$a_5$} &\sn0&\xn3&\sm0&\cn2&\sm0&\cn2&\sm0&\csn{blue}2&\sm0&\csn{blue}2&\csn{red}2&\sm0&\sn0&\sm0&\csn{red}2&\sm0&\sn0&\sm0&\csn{red}2&\csn{blue}2&\sm0&\csn{blue}2&\sm0&\cn2&\sm0&\cn2&\sm0&\xn3&\sn0& \\
\wire{$b_5$} &\sn0&\sn0&\sm0&\cn0&\sm0&\xn1&\sm0&\ccn{blue}1&\sm0&\csn{blue}2&\ccn{red}0&\sm0&\sn0&\sm0&\csn{red}2&\sm0&\sn0&\sm0&\ccn{red}0&\csn{blue}2&\sm0&\ccn{blue}1&\sm0&\xn1&\sm0&\cn0&\sm0&\sn0&\sn0& \\
\wire{$0$} &\sn0&\sn0&\sm0&\sn2&\sm0&\sn0&\sm0&\sn0&\sm0&\cxn{blue}0&\csn{red}2&\sm0&\sn0&\sm0&\ccn{red}0&\sm0&\sn0&\sm0&\csn{red}2&\cxn{blue}0&\sm0&\sn0&\sm0&\sn0&\sm0&\sn2&\sm0&\sn0&\sn0& \\
\wire{$0$} &\sn0&\sn0&\sm0&\xn1&\sm0&\sn0&\sm0&\sn0&\sm0&\csn{blue}2&\cxn{red}1&\sm0&\ccn{red}2&\sm0&\csn{red}2&\sm0&\sn0&\sm0&\cxn{red}1&\csn{blue}2&\sm0&\sn0&\sm0&\sn0&\sm0&\xn1&\sm0&\sn0&\sn0& \\
\wire{$a_6$} &\sn0&\xn3&\sm0&\cn2&\sm0&\cn2&\sm0&\sn0&\sm0&\csn{blue}2&\sn0&\sm0&\csn{red}2&\sm0&\csn{red}2&\sm0&\sn0&\sm0&\sn0&\csn{blue}2&\sm0&\sn0&\sm0&\cn2&\sm0&\sn0&\sm0&\xn3&\sn0& \\
\wire{$b_6$} &\sn0&\sn0&\sm0&\cn0&\sm0&\xn1&\sm0&\sn0&\sm0&\ccn{blue}1&\sn0&\sm0&\ccn{red}0&\sm0&\csn{red}2&\sm0&\sn0&\sm0&\sn0&\ccn{blue}1&\sm0&\sn0&\sm0&\xn1&\sm0&\sn0&\sm0&\sn0&\sn0& \\
\wire{$0$} &\sn0&\sn0&\sm0&\xn1&\sm0&\sn0&\sm0&\sn0&\sm0&\sn0&\sn0&\sm0&\cxn{red}1&\sm0&\cxn{red}1&\sm0&\sn0&\sm0&\sn0&\sn0&\sm0&\sn0&\sm0&\sn0&\sm0&\sn0&\sm0&\xn3&\sn0& \wire{$(b>a)?$} \\
\end{tabular}

\end{center}
\caption{\QCLA\ comparator for 7 bits.  $P$-rounds and
$P^{-1}$-rounds are shown in blue; $G$-rounds and $G^{-1}$-rounds are
red.}
\label{compare-fig}
\end{figure}

The total depth for the comparator is
$$
2\floor{\log (n-1)} + 9,
$$
where two of the time-slices contain negations, two contain
controlled-NOTs, and the rest are Toffolis.  The overall circuit size is
$$
6n - 2\floor{\log(n-1)} - w(n-1) - 7
$$
Toffoli gates, $2n - 2$ controlled-NOT gates, and $2n+1$ negations.
When $n \le 4$, we have slightly overcounted the depth and size.
A sample comparison circuit for $n = 7$ appears in Figure~\ref{compare-fig}.

If we wish to allow an incoming carry, we use the same technique as
in Section~\ref{incoming-carry-sec}.  We use an $(n+1)$-bit comparator,
except that the carry input replaces one of the ancillae, and we
save two negations and two Toffolis.

It may seem strange that an $n$-bit compare would require more
gates than an $n$-bit out-of-place add.  After all, we're solving
a simpler problem; we want one bit of the $(n+1)$-bit answer.

One way to look at this phenomenon is that, when we compute the high
bit of the sum, we are effectively using other output bits as ancillary
space.  If we're ``only'' doing a compare, we need extra gates to erase
this space.  One explicit example of this is
Step~\ref{compare-g-erase-step} of the compare, where we erase the
generate array.  For the out-of-place add, the generate array has turned
into our answer, and need not be erased.

\subsection{Addition \protect\boldmath{(mod $2^n - 1$)}}
\label{mod-mersenne-sec}
Recall from Section~\ref{notation-sec} that we have been working in
two's-complement arithmetic, where $r' + r = -1$.  With a slight
increase in depth, we can modify our circuit to work in one's-complement
arithmetic, where $r' + r = 0$.  Equivalently, we can view one's-complement
addition as addition (mod $2^n - 1$).  This may prove useful for
some applications, particularly when $2^n - 1$ is prime.

Note that, in one's-complement arithmetic, $0$ can be represented either
by the all-zeros bit string or the all-ones bit string.  For in-place
reversible computation, we cannot have $a + \vec{0} = a$ and
$a + \vec{1} = a$.  We will first describe our adder in general terms,
and then discuss how we can handle this zero problem.

First, consider the computation of $c_0$.  In the one's-complement
setting, we can no longer assume $c_0$ to be $0$; the low bit of the
sum is affected by whether or not we have an overflow.  We have an
overflow if and only if $a + b \ge 2^n$; hence, we get $c_0 = g[0,n]$.

If $n = 2^k$ for some $k$, then we have computed $c_0$ at the end of the
$G$-rounds.  How do we compute the other carry bits?  One approach
follows from the cyclic invariance of addition (mod $2^n - 1$):  We note
that multiplication by $2^j$ corresponds to a cyclic shift by $j$.
So, if we could simultaneously add at all possible cyclic shifts,
we would compute all of the carry bits.  This approach would have
logarithmic depth, but would require $\Theta(n \log n)$ ancillary space.

A second idea is to view $c_0$ as an incoming carry $g[-\infty, 0]$.
Our carry string is then given by $c_i = g[-\infty, i]$.  Another
way to look at this identity is that we are wrapping around:  to
compute $c_i$, we start at the zero position, work up to $n$, and
then wrap back around and keep going up to $i$.  This is the same
as the cyclic shift, except that we are counting one region twice; it
is not hard to see that this cannot affect our overall answer.

To do this wrap-around, we will need propagate bits of the form
$p[0,2^t]$. After we complete the $P$-rounds and $G$-rounds, we have
computed $c_0 = g[-\infty, 0]$.  We now add a new round: using $c_0$
and $p[0,2^{k-1}]$, we use one Toffoli gate to compute
$c_{2^{k-1}} = g[-\infty, 2^{k-1}]$.  From here on, we do our usual $C$-rounds,
except that each contains one extra gate computing $c_{2^t}$.  Upon
completion, we have successfully computed the carry string.

If $n$ is not a power of 2, we need to do a bit more work to make
sure we compute $c_0$ at the end of the $G$-rounds.  We use the
same technique as in Section~\ref{comparison-sec}.

\subsubsection{Out-of-place addition \protect\boldmath{(mod $2^n - 1$)}}
The above description, combined with the general approach in
Section~\ref{oop-qcla-sec}, yields an out-of-place one's-complement
adder.  We produce $n$ bits of output, and use $n-2$ ancillae.

The overall depth is
$$
2 \floor{\log(n-1)} + 8,
$$
where three of the time-slices contain controlled-NOTs and the
rest are Toffolis.  For $n \le 2$, the depth is slightly lower.

The circuit contains $5n - 6$ Toffoli gates and $3n$ controlled-NOT gates.

Suppose we use the above circuit to add two numbers $a$ and $b$ which
sum to $2^n - 1$.  Since $a \xor b = \vec{1}$, the generate array
will be initialized to $\vec{0}$ and the propagate array to $\vec{1}$.
Hence the carry string will be $\vec{0}$, and the sum will be output
as $\vec{1}$.  Hence, we say that this circuit uses the $\vec{1}$
representation of zero.  One can check that, if one of the inputs is
$\vec{1}$, the circuit also adds correctly.\footnote{In fact, this circuit
is also correct when exactly one of the inputs is $\vec{0}$.  But, if both inputs
are $\vec{0}$, we output $\vec{0}$ rather than $\vec{1}$.}

It might seem more natural to represent zero as $\vec{0}$.
We can modify the out-of-place circuit as follows:  at the end of the
$P$-rounds, we {\XOR} $p[0,n]$ into $c_0$ (this requires one additional
Toffoli gate).  So, after the $G$-rounds, we will have
$c_0 = g[0,n] \xor p[0,n]$. 
We then compute the $C$-rounds as before.  Now, if $a \xor b = \vec{1}$,
the circuit will output $\vec{0}$ as the sum; we have thus given a
circuit which uses the $\vec{0}$ representation of zero.  Again, we can
check that, if one or both inputs are $\vec{0}$, the circuit performs
correctly.

The out-of-place one's-complement adder using the $\vec{0}$ representation
requires $n-2$ ancillae.  The circuit contains $5n - 5$ Toffoli gates
and $3n$ controlled-NOT gates.  For some $n$, the depth goes up by one;
it depends on whether the computation of $p[0,n]$ can be done
simultaneously with the penultimate $G$-round.  For $n \ge 4$, the
depth can be written as
$$
\floor{\log(n-1)} + \floor{\log {n - 1\over 3}} + 10,
$$
where three of the time-slices contain controlled-NOTs and the
rest are Toffolis.

\subsubsection{In-place addition \protect\boldmath{(mod $2^n - 1$)}}

Following Section~\ref{ip-qcla-sec}, we next construct
an in-place one's-complement adder.  We require $2n - 2$ ancillary
bits:  $n-2$ for computing the propagate bits, and $n$ for the
carry string.  We first compute the carry string into our ancillae,
and then write the sum on top of $b$.  Next, to erase the carry
string, we negate $b$, undo the addition computation, and fix $b$
at the end.

\begin{sidewaysfigure}[p]
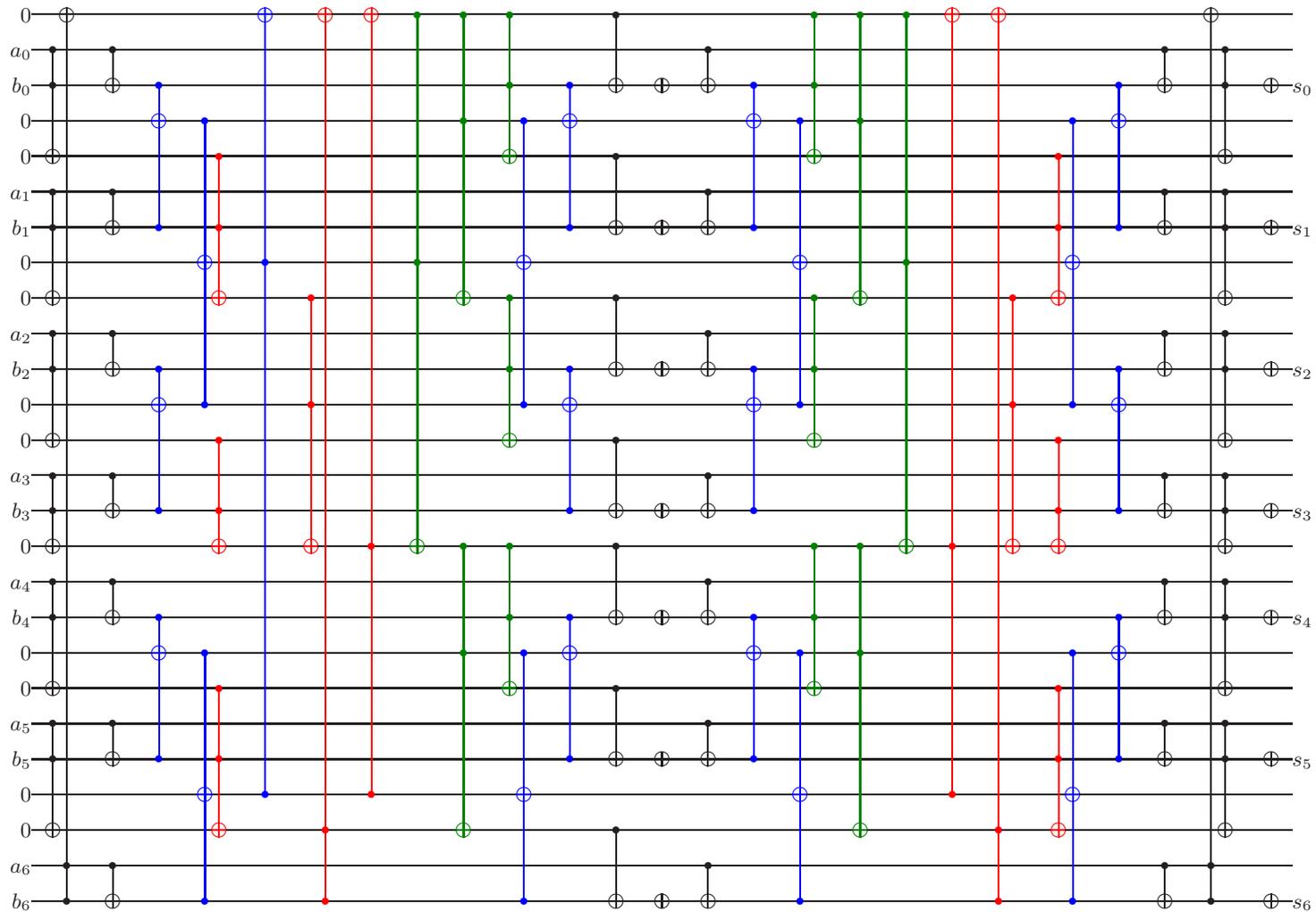

\renewcommand{\arraystretch}{0}
\renewcommand{\smwidth}{9}
\renewcommand{\smhalf}{4}
\begin{tabular}{r@{}*{59}{c@{}}l}
\wire{$0$} &\sn0&\sn0&\xn2&\sm0&\sn0&\sm0&\sn0&\sm0&\sn0&\sn0&\sm0&\cxn{blue}2&\sm0&\sn0&\cxn{red}2&\sm0&\cxn{red}2&\sm0&\ccn{darkgreen}2&\sm0&\ccn{darkgreen}2&\sm0&\ccn{darkgreen}2&\sn0&\sm0&\sn0&\sm0&\cn2&\sm0&\sn0&\sm0&\sn0&\sm0&\sn0&\sm0&\sn0&\ccn{darkgreen}2&\sm0&\ccn{darkgreen}2&\sm0&\ccn{darkgreen}2&\sm0&\cxn{red}2&\sm0&\cxn{red}2&\sn0&\sm0&\sn0&\sn0&\sm0&\sn0&\sm0&\sn0&\sm0&\xn2&\sn0&\sm0&\sn0&\sn0& \\
\wire{$a_0$} &\sn0&\cn2&\sn2&\sm0&\cn2&\sm0&\sn0&\sm0&\sn0&\sn0&\sm0&\csn{blue}2&\sm0&\sn0&\csn{red}2&\sm0&\csn{red}2&\sm0&\csn{darkgreen}2&\sm0&\csn{darkgreen}2&\sm0&\csn{darkgreen}2&\sn0&\sm0&\sn0&\sm0&\sn2&\sm0&\sn0&\sm0&\cn2&\sm0&\sn0&\sm0&\sn0&\csn{darkgreen}2&\sm0&\csn{darkgreen}2&\sm0&\csn{darkgreen}2&\sm0&\csn{red}2&\sm0&\csn{red}2&\sn0&\sm0&\sn0&\sn0&\sm0&\sn0&\sm0&\cn2&\sm0&\sn2&\cn2&\sm0&\sn0&\sn0& \\
\wire{$b_0$} &\sn0&\cn0&\sn2&\sm0&\xn1&\sm0&\ccn{blue}2&\sm0&\sn0&\sn0&\sm0&\csn{blue}2&\sm0&\sn0&\csn{red}2&\sm0&\csn{red}2&\sm0&\csn{darkgreen}2&\sm0&\csn{darkgreen}2&\sm0&\ccn{darkgreen}0&\sn0&\sm0&\ccn{blue}2&\sm0&\xn1&\sm0&\xn3&\sm0&\xn1&\sm0&\ccn{blue}2&\sm0&\sn0&\ccn{darkgreen}0&\sm0&\csn{darkgreen}2&\sm0&\csn{darkgreen}2&\sm0&\csn{red}2&\sm0&\csn{red}2&\sn0&\sm0&\sn0&\sn0&\sm0&\ccn{blue}2&\sm0&\xn1&\sm0&\sn2&\cn0&\sm0&\xn3&\sn0& \wire{$s_0$} \\
\wire{$0$} &\sn0&\sn2&\sn2&\sm0&\sn0&\sm0&\cxn{blue}0&\sm0&\ccn{blue}2&\sn0&\sm0&\csn{blue}2&\sm0&\sn0&\csn{red}2&\sm0&\csn{red}2&\sm0&\csn{darkgreen}2&\sm0&\ccn{darkgreen}0&\sm0&\csn{darkgreen}2&\ccn{blue}2&\sm0&\cxn{blue}0&\sm0&\sn0&\sm0&\sn0&\sm0&\sn0&\sm0&\cxn{blue}0&\sm0&\ccn{blue}2&\csn{darkgreen}2&\sm0&\ccn{darkgreen}0&\sm0&\csn{darkgreen}2&\sm0&\csn{red}2&\sm0&\csn{red}2&\sn0&\sm0&\sn0&\ccn{blue}2&\sm0&\cxn{blue}0&\sm0&\sn0&\sm0&\sn2&\sn2&\sm0&\sn0&\sn0& \\
\wire{$0$} &\sn0&\xn1&\sn2&\sm0&\sn0&\sm0&\csn{blue}2&\sm0&\csn{blue}2&\ccn{red}2&\sm0&\csn{blue}2&\sm0&\sn0&\csn{red}2&\sm0&\csn{red}2&\sm0&\csn{darkgreen}2&\sm0&\csn{darkgreen}2&\sm0&\cxn{darkgreen}1&\csn{blue}2&\sm0&\csn{blue}2&\sm0&\cn2&\sm0&\sn0&\sm0&\sn0&\sm0&\csn{blue}2&\sm0&\csn{blue}2&\cxn{darkgreen}1&\sm0&\csn{darkgreen}2&\sm0&\csn{darkgreen}2&\sm0&\csn{red}2&\sm0&\csn{red}2&\sn0&\sm0&\ccn{red}2&\csn{blue}2&\sm0&\csn{blue}2&\sm0&\sn0&\sm0&\sn2&\xn1&\sm0&\sn0&\sn0& \\
\wire{$a_1$} &\sn0&\cn2&\sn2&\sm0&\cn2&\sm0&\csn{blue}2&\sm0&\csn{blue}2&\csn{red}2&\sm0&\csn{blue}2&\sm0&\sn0&\csn{red}2&\sm0&\csn{red}2&\sm0&\csn{darkgreen}2&\sm0&\csn{darkgreen}2&\sm0&\sn0&\csn{blue}2&\sm0&\csn{blue}2&\sm0&\sn2&\sm0&\sn0&\sm0&\cn2&\sm0&\csn{blue}2&\sm0&\csn{blue}2&\sn0&\sm0&\csn{darkgreen}2&\sm0&\csn{darkgreen}2&\sm0&\csn{red}2&\sm0&\csn{red}2&\sn0&\sm0&\csn{red}2&\csn{blue}2&\sm0&\csn{blue}2&\sm0&\cn2&\sm0&\sn2&\cn2&\sm0&\sn0&\sn0& \\
\wire{$b_1$} &\sn0&\cn0&\sn2&\sm0&\xn1&\sm0&\ccn{blue}1&\sm0&\csn{blue}2&\ccn{red}0&\sm0&\csn{blue}2&\sm0&\sn0&\csn{red}2&\sm0&\csn{red}2&\sm0&\csn{darkgreen}2&\sm0&\csn{darkgreen}2&\sm0&\sn0&\csn{blue}2&\sm0&\ccn{blue}1&\sm0&\xn1&\sm0&\xn3&\sm0&\xn1&\sm0&\ccn{blue}1&\sm0&\csn{blue}2&\sn0&\sm0&\csn{darkgreen}2&\sm0&\csn{darkgreen}2&\sm0&\csn{red}2&\sm0&\csn{red}2&\sn0&\sm0&\ccn{red}0&\csn{blue}2&\sm0&\ccn{blue}1&\sm0&\xn1&\sm0&\sn2&\cn0&\sm0&\xn3&\sn0& \wire{$s_1$} \\
\wire{$0$} &\sn0&\sn2&\sn2&\sm0&\sn0&\sm0&\sn0&\sm0&\cxn{blue}0&\csn{red}2&\sm0&\ccn{blue}0&\sm0&\sn0&\csn{red}2&\sm0&\csn{red}2&\sm0&\ccn{darkgreen}0&\sm0&\csn{darkgreen}2&\sm0&\sn0&\cxn{blue}0&\sm0&\sn0&\sm0&\sn0&\sm0&\sn0&\sm0&\sn0&\sm0&\sn0&\sm0&\cxn{blue}0&\sn0&\sm0&\csn{darkgreen}2&\sm0&\ccn{darkgreen}0&\sm0&\csn{red}2&\sm0&\csn{red}2&\sn0&\sm0&\csn{red}2&\cxn{blue}0&\sm0&\sn0&\sm0&\sn0&\sm0&\sn2&\sn2&\sm0&\sn0&\sn0& \\
\wire{$0$} &\sn0&\xn1&\sn2&\sm0&\sn0&\sm0&\sn0&\sm0&\csn{blue}2&\cxn{red}1&\sm0&\csn{blue}2&\sm0&\ccn{red}2&\csn{red}2&\sm0&\csn{red}2&\sm0&\csn{darkgreen}2&\sm0&\cxn{darkgreen}1&\sm0&\ccn{darkgreen}2&\csn{blue}2&\sm0&\sn0&\sm0&\cn2&\sm0&\sn0&\sm0&\sn0&\sm0&\sn0&\sm0&\csn{blue}2&\ccn{darkgreen}2&\sm0&\cxn{darkgreen}1&\sm0&\csn{darkgreen}2&\sm0&\csn{red}2&\sm0&\csn{red}2&\ccn{red}2&\sm0&\cxn{red}1&\csn{blue}2&\sm0&\sn0&\sm0&\sn0&\sm0&\sn2&\xn1&\sm0&\sn0&\sn0& \\
\wire{$a_2$} &\sn0&\cn2&\sn2&\sm0&\cn2&\sm0&\sn0&\sm0&\csn{blue}2&\sn0&\sm0&\csn{blue}2&\sm0&\csn{red}2&\csn{red}2&\sm0&\csn{red}2&\sm0&\csn{darkgreen}2&\sm0&\sn0&\sm0&\csn{darkgreen}2&\csn{blue}2&\sm0&\sn0&\sm0&\sn2&\sm0&\sn0&\sm0&\cn2&\sm0&\sn0&\sm0&\csn{blue}2&\csn{darkgreen}2&\sm0&\sn0&\sm0&\csn{darkgreen}2&\sm0&\csn{red}2&\sm0&\csn{red}2&\csn{red}2&\sm0&\sn0&\csn{blue}2&\sm0&\sn0&\sm0&\cn2&\sm0&\sn2&\cn2&\sm0&\sn0&\sn0& \\
\wire{$b_2$} &\sn0&\cn0&\sn2&\sm0&\xn1&\sm0&\ccn{blue}2&\sm0&\csn{blue}2&\sn0&\sm0&\csn{blue}2&\sm0&\csn{red}2&\csn{red}2&\sm0&\csn{red}2&\sm0&\csn{darkgreen}2&\sm0&\sn0&\sm0&\ccn{darkgreen}0&\csn{blue}2&\sm0&\ccn{blue}2&\sm0&\xn1&\sm0&\xn3&\sm0&\xn1&\sm0&\ccn{blue}2&\sm0&\csn{blue}2&\ccn{darkgreen}0&\sm0&\sn0&\sm0&\csn{darkgreen}2&\sm0&\csn{red}2&\sm0&\csn{red}2&\csn{red}2&\sm0&\sn0&\csn{blue}2&\sm0&\ccn{blue}2&\sm0&\xn1&\sm0&\sn2&\cn0&\sm0&\xn3&\sn0& \wire{$s_2$} \\
\wire{$0$} &\sn0&\sn2&\sn2&\sm0&\sn0&\sm0&\cxn{blue}0&\sm0&\ccn{blue}1&\sn0&\sm0&\csn{blue}2&\sm0&\ccn{red}0&\csn{red}2&\sm0&\csn{red}2&\sm0&\csn{darkgreen}2&\sm0&\sn0&\sm0&\csn{darkgreen}2&\ccn{blue}1&\sm0&\cxn{blue}0&\sm0&\sn0&\sm0&\sn0&\sm0&\sn0&\sm0&\cxn{blue}0&\sm0&\ccn{blue}1&\csn{darkgreen}2&\sm0&\sn0&\sm0&\csn{darkgreen}2&\sm0&\csn{red}2&\sm0&\csn{red}2&\ccn{red}0&\sm0&\sn0&\ccn{blue}1&\sm0&\cxn{blue}0&\sm0&\sn0&\sm0&\sn2&\sn2&\sm0&\sn0&\sn0& \\
\wire{$0$} &\sn0&\xn1&\sn2&\sm0&\sn0&\sm0&\csn{blue}2&\sm0&\sn0&\ccn{red}2&\sm0&\csn{blue}2&\sm0&\csn{red}2&\csn{red}2&\sm0&\csn{red}2&\sm0&\csn{darkgreen}2&\sm0&\sn0&\sm0&\cxn{darkgreen}1&\sn0&\sm0&\csn{blue}2&\sm0&\cn2&\sm0&\sn0&\sm0&\sn0&\sm0&\csn{blue}2&\sm0&\sn0&\cxn{darkgreen}1&\sm0&\sn0&\sm0&\csn{darkgreen}2&\sm0&\csn{red}2&\sm0&\csn{red}2&\csn{red}2&\sm0&\ccn{red}2&\sn0&\sm0&\csn{blue}2&\sm0&\sn0&\sm0&\sn2&\xn1&\sm0&\sn0&\sn0& \\
\wire{$a_3$} &\sn0&\cn2&\sn2&\sm0&\cn2&\sm0&\csn{blue}2&\sm0&\sn0&\csn{red}2&\sm0&\csn{blue}2&\sm0&\csn{red}2&\csn{red}2&\sm0&\csn{red}2&\sm0&\csn{darkgreen}2&\sm0&\sn0&\sm0&\sn0&\sn0&\sm0&\csn{blue}2&\sm0&\sn2&\sm0&\sn0&\sm0&\cn2&\sm0&\csn{blue}2&\sm0&\sn0&\sn0&\sm0&\sn0&\sm0&\csn{darkgreen}2&\sm0&\csn{red}2&\sm0&\csn{red}2&\csn{red}2&\sm0&\csn{red}2&\sn0&\sm0&\csn{blue}2&\sm0&\cn2&\sm0&\sn2&\cn2&\sm0&\sn0&\sn0& \\
\wire{$b_3$} &\sn0&\cn0&\sn2&\sm0&\xn1&\sm0&\ccn{blue}1&\sm0&\sn0&\ccn{red}0&\sm0&\csn{blue}2&\sm0&\csn{red}2&\csn{red}2&\sm0&\csn{red}2&\sm0&\csn{darkgreen}2&\sm0&\sn0&\sm0&\sn0&\sn0&\sm0&\ccn{blue}1&\sm0&\xn1&\sm0&\xn3&\sm0&\xn1&\sm0&\ccn{blue}1&\sm0&\sn0&\sn0&\sm0&\sn0&\sm0&\csn{darkgreen}2&\sm0&\csn{red}2&\sm0&\csn{red}2&\csn{red}2&\sm0&\ccn{red}0&\sn0&\sm0&\ccn{blue}1&\sm0&\xn1&\sm0&\sn2&\cn0&\sm0&\xn3&\sn0& \wire{$s_3$} \\
\wire{$0$} &\sn0&\xn1&\sn2&\sm0&\sn0&\sm0&\sn0&\sm0&\sn0&\cxn{red}1&\sm0&\csn{blue}2&\sm0&\cxn{red}1&\csn{red}2&\sm0&\ccn{red}0&\sm0&\cxn{darkgreen}1&\sm0&\ccn{darkgreen}2&\sm0&\ccn{darkgreen}2&\sn0&\sm0&\sn0&\sm0&\cn2&\sm0&\sn0&\sm0&\sn0&\sm0&\sn0&\sm0&\sn0&\ccn{darkgreen}2&\sm0&\ccn{darkgreen}2&\sm0&\cxn{darkgreen}1&\sm0&\ccn{red}0&\sm0&\csn{red}2&\cxn{red}1&\sm0&\cxn{red}1&\sn0&\sm0&\sn0&\sm0&\sn0&\sm0&\sn2&\xn1&\sm0&\sn0&\sn0& \\
\wire{$a_4$} &\sn0&\cn2&\sn2&\sm0&\cn2&\sm0&\sn0&\sm0&\sn0&\sn0&\sm0&\csn{blue}2&\sm0&\sn0&\csn{red}2&\sm0&\csn{red}2&\sm0&\sn0&\sm0&\csn{darkgreen}2&\sm0&\csn{darkgreen}2&\sn0&\sm0&\sn0&\sm0&\sn2&\sm0&\sn0&\sm0&\cn2&\sm0&\sn0&\sm0&\sn0&\csn{darkgreen}2&\sm0&\csn{darkgreen}2&\sm0&\sn0&\sm0&\csn{red}2&\sm0&\csn{red}2&\sn0&\sm0&\sn0&\sn0&\sm0&\sn0&\sm0&\cn2&\sm0&\sn2&\cn2&\sm0&\sn0&\sn0& \\
\wire{$b_4$} &\sn0&\cn0&\sn2&\sm0&\xn1&\sm0&\ccn{blue}2&\sm0&\sn0&\sn0&\sm0&\csn{blue}2&\sm0&\sn0&\csn{red}2&\sm0&\csn{red}2&\sm0&\sn0&\sm0&\csn{darkgreen}2&\sm0&\ccn{darkgreen}0&\sn0&\sm0&\ccn{blue}2&\sm0&\xn1&\sm0&\xn3&\sm0&\xn1&\sm0&\ccn{blue}2&\sm0&\sn0&\ccn{darkgreen}0&\sm0&\csn{darkgreen}2&\sm0&\sn0&\sm0&\csn{red}2&\sm0&\csn{red}2&\sn0&\sm0&\sn0&\sn0&\sm0&\ccn{blue}2&\sm0&\xn1&\sm0&\sn2&\cn0&\sm0&\xn3&\sn0& \wire{$s_4$} \\
\wire{$0$} &\sn0&\sn2&\sn2&\sm0&\sn0&\sm0&\cxn{blue}0&\sm0&\ccn{blue}2&\sn0&\sm0&\csn{blue}2&\sm0&\sn0&\csn{red}2&\sm0&\csn{red}2&\sm0&\sn0&\sm0&\ccn{darkgreen}0&\sm0&\csn{darkgreen}2&\ccn{blue}2&\sm0&\cxn{blue}0&\sm0&\sn0&\sm0&\sn0&\sm0&\sn0&\sm0&\cxn{blue}0&\sm0&\ccn{blue}2&\csn{darkgreen}2&\sm0&\ccn{darkgreen}0&\sm0&\sn0&\sm0&\csn{red}2&\sm0&\csn{red}2&\sn0&\sm0&\sn0&\ccn{blue}2&\sm0&\cxn{blue}0&\sm0&\sn0&\sm0&\sn2&\sn2&\sm0&\sn0&\sn0& \\
\wire{$0$} &\sn0&\xn1&\sn2&\sm0&\sn0&\sm0&\csn{blue}2&\sm0&\csn{blue}2&\ccn{red}2&\sm0&\csn{blue}2&\sm0&\sn0&\csn{red}2&\sm0&\csn{red}2&\sm0&\sn0&\sm0&\csn{darkgreen}2&\sm0&\cxn{darkgreen}1&\csn{blue}2&\sm0&\csn{blue}2&\sm0&\cn2&\sm0&\sn0&\sm0&\sn0&\sm0&\csn{blue}2&\sm0&\csn{blue}2&\cxn{darkgreen}1&\sm0&\csn{darkgreen}2&\sm0&\sn0&\sm0&\csn{red}2&\sm0&\csn{red}2&\sn0&\sm0&\ccn{red}2&\csn{blue}2&\sm0&\csn{blue}2&\sm0&\sn0&\sm0&\sn2&\xn1&\sm0&\sn0&\sn0& \\
\wire{$a_5$} &\sn0&\cn2&\sn2&\sm0&\cn2&\sm0&\csn{blue}2&\sm0&\csn{blue}2&\csn{red}2&\sm0&\csn{blue}2&\sm0&\sn0&\csn{red}2&\sm0&\csn{red}2&\sm0&\sn0&\sm0&\csn{darkgreen}2&\sm0&\sn0&\csn{blue}2&\sm0&\csn{blue}2&\sm0&\sn2&\sm0&\sn0&\sm0&\cn2&\sm0&\csn{blue}2&\sm0&\csn{blue}2&\sn0&\sm0&\csn{darkgreen}2&\sm0&\sn0&\sm0&\csn{red}2&\sm0&\csn{red}2&\sn0&\sm0&\csn{red}2&\csn{blue}2&\sm0&\csn{blue}2&\sm0&\cn2&\sm0&\sn2&\cn2&\sm0&\sn0&\sn0& \\
\wire{$b_5$} &\sn0&\cn0&\sn2&\sm0&\xn1&\sm0&\ccn{blue}1&\sm0&\csn{blue}2&\ccn{red}0&\sm0&\csn{blue}2&\sm0&\sn0&\csn{red}2&\sm0&\csn{red}2&\sm0&\sn0&\sm0&\csn{darkgreen}2&\sm0&\sn0&\csn{blue}2&\sm0&\ccn{blue}1&\sm0&\xn1&\sm0&\xn3&\sm0&\xn1&\sm0&\ccn{blue}1&\sm0&\csn{blue}2&\sn0&\sm0&\csn{darkgreen}2&\sm0&\sn0&\sm0&\csn{red}2&\sm0&\csn{red}2&\sn0&\sm0&\ccn{red}0&\csn{blue}2&\sm0&\ccn{blue}1&\sm0&\xn1&\sm0&\sn2&\cn0&\sm0&\xn3&\sn0& \wire{$s_5$} \\
\wire{$0$} &\sn0&\sn2&\sn2&\sm0&\sn0&\sm0&\sn0&\sm0&\cxn{blue}0&\csn{red}2&\sm0&\ccn{blue}1&\sm0&\sn0&\csn{red}2&\sm0&\ccn{red}1&\sm0&\sn0&\sm0&\csn{darkgreen}2&\sm0&\sn0&\cxn{blue}0&\sm0&\sn0&\sm0&\sn0&\sm0&\sn0&\sm0&\sn0&\sm0&\sn0&\sm0&\cxn{blue}0&\sn0&\sm0&\csn{darkgreen}2&\sm0&\sn0&\sm0&\ccn{red}1&\sm0&\csn{red}2&\sn0&\sm0&\csn{red}2&\cxn{blue}0&\sm0&\sn0&\sm0&\sn0&\sm0&\sn2&\sn2&\sm0&\sn0&\sn0& \\
\wire{$0$} &\sn0&\xn1&\sn2&\sm0&\sn0&\sm0&\sn0&\sm0&\csn{blue}2&\cxn{red}1&\sm0&\sn0&\sm0&\sn0&\ccn{red}0&\sm0&\sn0&\sm0&\sn0&\sm0&\cxn{darkgreen}1&\sm0&\sn0&\csn{blue}2&\sm0&\sn0&\sm0&\cn2&\sm0&\sn0&\sm0&\sn0&\sm0&\sn0&\sm0&\csn{blue}2&\sn0&\sm0&\cxn{darkgreen}1&\sm0&\sn0&\sm0&\sn0&\sm0&\ccn{red}0&\sn0&\sm0&\cxn{red}1&\csn{blue}2&\sm0&\sn0&\sm0&\sn0&\sm0&\sn2&\xn1&\sm0&\sn0&\sn0& \\
\wire{$a_6$} &\sn0&\sn0&\cn0&\sm0&\cn2&\sm0&\sn0&\sm0&\csn{blue}2&\sn0&\sm0&\sn0&\sm0&\sn0&\csn{red}2&\sm0&\sn0&\sm0&\sn0&\sm0&\sn0&\sm0&\sn0&\csn{blue}2&\sm0&\sn0&\sm0&\sn2&\sm0&\sn0&\sm0&\cn2&\sm0&\sn0&\sm0&\csn{blue}2&\sn0&\sm0&\sn0&\sm0&\sn0&\sm0&\sn0&\sm0&\csn{red}2&\sn0&\sm0&\sn0&\csn{blue}2&\sm0&\sn0&\sm0&\cn2&\sm0&\cn0&\sn0&\sm0&\sn0&\sn0& \\
\wire{$b_6$} &\sn0&\sn0&\cn1&\sm0&\xn1&\sm0&\sn0&\sm0&\ccn{blue}1&\sn0&\sm0&\sn0&\sm0&\sn0&\ccn{red}1&\sm0&\sn0&\sm0&\sn0&\sm0&\sn0&\sm0&\sn0&\ccn{blue}1&\sm0&\sn0&\sm0&\xn1&\sm0&\xn3&\sm0&\xn1&\sm0&\sn0&\sm0&\ccn{blue}1&\sn0&\sm0&\sn0&\sm0&\sn0&\sm0&\sn0&\sm0&\ccn{red}1&\sn0&\sm0&\sn0&\ccn{blue}1&\sm0&\sn0&\sm0&\xn1&\sm0&\cn1&\sn0&\sm0&\xn3&\sn0& \wire{$s_6$} \\
\end{tabular}

\caption{In-place \QCLA\ adder (mod $2^n - 1$) for $n = 7$.    $P$-rounds and
$P^{-1}$-rounds are shown in blue.  $G$-rounds are red, and $C$-rounds are
green.
This adder uses the $\vec{0}$ representation of zero.  Note that the extra
gate writing $p[0,n]$ to $c_0$ is present only during the first half of
the computation.}
\label{ip-mersenne-fig}
\end{sidewaysfigure}

As in the out-of-place version, we need to be careful about the
representation of zero.  The complementation of $b$ introduces
a slight wrinkle:  if we do our first addition using $\vec{0}$ to
represent zero, then we need to undo a $\vec{1}$-based addition.
If we do our first addition using the $\vec{1}$
circuit, we need to undo the $\vec{0}$ circuit.

Hence, regardless of whether we represent zero by $\vec{0}$ or
$\vec{1}$, the cost of the circuit is the same:  we require
$2n$ negations, $4n$ controlled-NOTs, and $10n-11$ Toffolis.  For
$n \ge 4$, the depth is:
$$
3\floor{\log(n-1)} + \floor{\log {n - 1\over 3}} + 18,
$$
where two of the time-slices contain negations, four contain
controlled-NOTs, and the rest contain Toffolis.

Figure~\ref{ip-mersenne-fig} depicts a sample in-place one's-complement
\QCLA\ adder for the case $n=7$.

\section{Conclusions and future work}

In conclusion, we have developed an efficient addition circuit using
classical carry-lookahead techniques.  Our \QCLA\ adder sums two
$n$-bit numbers in-place using $2n - w(n) - \floor{\log n} - 1$
ancillary qubits in depth $4 \log n + O(1)$.  This improves upon the
previous best known addition circuits, which require linear depth.
Our work dramatically improves the run-time of the arithmetic circuits
required in Shor's algorithm.

\begin{table}
\begin{center}
\renewcommand{\arraystretch}{1.5}
\begin{tabular}{|c|cc||ccc|c|c|}\hline
Function & IP? & IC? & Input & Output & Ancillae & Size & Depth \\ \hline \hline
$+$ in $\Z$ & N & N & $2n$ & $n+1$ & $n - k - 1$ &
$5n - 3k - 4$ & $2k + 2$ \\
$+$ in $\Z$ & N & Y & $2n+1$ & $n+1$ & $n - k - 1$ &
$5n - 3k - 3$ & $2k + 2$ \\ \hline
$+$ in $\Z$ & Y & N & $2n$ & 1 & $2n - k - 2$ &
$10n - 9k - 7$ & $4k + 3$ \\
$+$ in $\Z$ & Y & Y & $2n+1$ & 1 & $2n - k - 2$ &
$10n - 6k - 8$ & $4k + 4$\\ \hline
$+$ (mod $2^n$) & N & N & $2n$ & $n$ & $n - 2k$ &
$5n - 6k - 3$ & $2k + 1$ \\
$+$ (mod $2^n$) & N & Y & $2n+1$ & $n$ & $n - k - 1$ &
$5n - 3k - 5$ & $2k + 2$ \\ \hline
$+$ (mod $2^n$) & Y & N & $2n$ & 0 & $2n - 2k - 1$ &
$10n - 12k - 6$ & $4k + 2$ \\
$+$ (mod $2^n$) & Y & Y & $2n+1$ & 0 & $2n - k - 2$ &
$10n - 6k - 10$ & $4k + 4$ \\ \hline
$+$ (mod $2^n - 1$) & N & --- & $2n$ & $n$ & $n-2$ & $5n-6$ & $2k+3$ \\ \hline
$+$ (mod $2^n - 1$) & Y & --- & $2n$ & 0 & $2n-2$ & $10n-11$ & $4k+7$ \\ \hline
Comparison & --- & N & $2n$ & 1 & $2n - k - 2$ &
$6n - 3k - 5$ & $2k + 5$ \\ 
Comparison & --- & Y & $2n + 1$ & 1 & $2n - k - 2$ &
$6n - 2k - 4$ & $2k + 5$ \\ \hline
Ripple-carry~\cite{VBE} & Y & N & $2n$ & 1 & $n$ & $4n - 2$ & $3n - 1$ \\
Ripple-carry~\cite{ripple-carry-add} & Y & Y & $2n+1$ & 1 & 0 & 
$2n - 1$ & $2n - 1$ \\ \hline
\end{tabular}
\end{center}
\caption{Circuit summary, for $n = 2^k$, where $k \ge 3$.
The first column gives the function being
computed.  The second lists whether the computation is done in place, and the
third lists whether we take an incoming carry bit as input.  We then
list the number of input, output, and ancillae bits, and the circuit
size and depth.  For the purposes of this table, we only count Toffoli
gates.}
\label{summary-table}
\end{table}

\begin{sidewaystable}
\begin{center}
\scriptsize
\renewcommand{\arraystretch}{1.5}
\begin{tabular}{|c|cc||ccc|c|c|}\hline
Function & IP? & IC? & Input & Output & Ancillae & Size & Depth \\ \hline \hline
$+$ in $\Z$ & N & N & $2n$ & $n+1$ & $n - w(n) - \floor{\log n}$ &
$5n - 3w(n) - 3 \floor{\log n} - 1$ &
$\floor{\log n} + \floor{\log {n \over 3}} + 4$ \\
$+$ in $\Z$ & N & Y & $2n+1$ & $n+1$ & $n - w(n+1) - \floor{\log(n+1)} + 1$ &
$5n - 3w(n+1) - 3 \floor{\log (n+1)} + 3$ &
$\floor{\log(n+1)} + \floor{\log {n + 1\over 3}} + 4$ \\ \hline
$+$ in $\Z$ & Y & N & $2n$ & 1 & $2n - w(n) - \floor{\log n} - 1$ &
\multicolumn{1}{l|}{$10n - 3w(n) - 3w(n-1)$}&
\multicolumn{1}{l|}{$\floor{\log n} + \floor{\log(n-1)} + \floor{\log{n \over 3}}$} \\
&&&&&&\multicolumn{1}{r|}{${} - 3 \floor{\log n} - 3 \floor{\log(n-1)} - 7$}&
\multicolumn{1}{r|}{${} + \floor{\log{n - 1 \over 3}} + 8$}\\
$+$ in $\Z$ & Y & Y & $2n+1$ & 1 & $2n - w(n+1) - \floor{\log(n+1)}$ &
\multicolumn{1}{l|}{$10n - 3w(n) - 3w(n+1)$} &
\multicolumn{1}{l|}{$\floor{\log n} + \floor{\log(n+1)} + \floor{\log{n \over 3}}$}\\
&&&&&&\multicolumn{1}{r|}{${} - 3 \floor{\log n} - 3 \floor{\log(n+1)} + 1$} &
\multicolumn{1}{r|}{$ + \floor{\log{n + 1 \over 3}} + 8$} \\ \hline
$+$ (mod $2^n$) & N & N & $2n$ & $n$ & $n - w(n-1) - \floor{\log(n-1)} - 1$ &
$5n - 3w(n-1) - 3 \floor{\log (n-1)} - 6$ &
$\floor{\log (n-1)} + \floor{\log{n-1\over 3}} + 4$ \\
$+$ (mod $2^n$) & N & Y & $2n+1$ & $n$ & $n - w(n) - \floor{\log n}$ &
$5n - 3w(n) - 3 \floor{\log n} - 2$ &
$\floor{\log n} + \floor{\log{n\over 3}} + 4$ \\ \hline
$+$ (mod $2^n$) & Y & N & $2n$ & 0 & $2n - w(n-1) - \floor{\log(n-1)} - 2$ &
$10n - 6w(n-1) - 6 \floor{\log(n-1)} - 12$ &
$2\floor{\log(n-1)} + 2 \floor{\log{n-1 \over 3}} + 8$ \\
$+$ (mod $2^n$) & Y & Y & $2n+1$ & 0 & $2n - w(n) - \floor{\log n} - 1$ &
$10n - 6w(n) - 6 \floor{\log n} - 4$ &
$2\floor{\log n} + 2 \floor{\log{n \over 3}} + 8$ \\ \hline
$+$ (mod $2^n - 1$) & N & --- & $2n$ & $n$ & $n-2$ & $5n - 6$ & $2\floor{\log(n-1)} + 5$ \\ \hline
$+$ (mod $2^n - 1$) & Y & --- & $2n$ & 0 & $2n-2$ & $10n - 11$ & $3\floor{\log(n-1)} + \floor{\log{n-1 \over 3}} + 12$ \\ \hline
Comparison & --- & N & $2n$ & 1 & $2n - \floor{\log(n-1)} - 3$ &
$6n - w(n-1) - 2 \floor{\log(n-1)} - 7$ & $2 \floor{\log n} + 5$ \\
Comparison & --- & Y & $2n+1$ & 1 & $2n - \floor{\log n} - 2$ &
$6n - w(n) - 2 \floor{\log n} - 3$ & $2 \floor{\log (n+1)} + 5$ \\ \hline
Ripple-carry~\cite{VBE} & Y & N & $2n$ & 1 & $n$ & $4n - 2$ & $3n - 1$ \\
Ripple-carry~\cite{ripple-carry-add} & Y & Y & $2n+1$ & 1 & 0 & 
$2n - 1$ & $2n - 1$ \\ \hline
\end{tabular}
\end{center}
\caption{Circuit summary, for $n \ge 7$.
The first column gives the function being
computed.  The second lists whether the computation is done in place, and the
third lists whether we take an incoming carry bit as input.  We then
list the number of input, output, and ancillae bits, and the circuit
size and depth.  For the purposes of this table, we only count Toffoli
gates.  Recall that $w(n)$ is the number of ones in the binary expansion of $n$.}
\label{summary-table-big}
\end{sidewaystable}

The complexities of the various circuits in this paper are summarized
in Tables~\ref{summary-table} and~\ref{summary-table-big}.  In
Table~\ref{summary-table}, we assume $n = 2^k$; in
Table~\ref{summary-table-big}, we give the general formulas.
For simplicity, we count only Toffoli gates and Toffoli time-slices.
Since some of the formulas are incorrect for small $n$, we assume $n \ge 7$.
We also include two different ripple-carry adders~\cite{VBE,ripple-carry-add}.

It would be interesting to apply a similar tree-like approach
to other arithmetic problems, such as modular addition and
multiplication.
It would also be interesting to build a logarithmic-depth addition circuit
using only $O(\log n)$ ancillae, or to prove that no such classical
reversible circuit exists.

\bibliography{adder}
\bibliographystyle{amsplain}

\end{document}